# Assessing User Interface Design Artifacts: A Tool-Supported Behavior-Based Approach*

Thiago Rocha Silva[1], Marco Winckler[2]

[1] *The Maersk Mc-Kinney Moller Institute, University of Southern Denmark, Odense, Denmark*

[2] *SPARKS-i3S, Université Nice Sophia Antipolis (Polytech), Sophia Antipolis, France*

The Maersk Mc-Kinney Moller Institute, University of Southern Denmark, Odense, Denmark.

Corresponding author: Thiago Rocha Silva (thiago@mmmi.sdu.dk)

---

* This article is an extended and revised version of the short paper entitled "*Extending Behavior-Driven Development for Assessing User Interface Design Artifacts*" published in SEKE 2019.

# Assessing User Interface Design Artifacts: A Tool-Supported Behavior-Based Approach

Thiago Rocha Silva
The Maersk Mc-Kinney Moller Institute
University of Southern Denmark
Odense, Denmark
thiago@mmmi.sdu.dk

Marco Winckler
SPARKS-i3S
Université Nice Sophia Antipolis (Polytech)
Sophia Antipolis, France
winckler@unice.fr

**Abstract:** Behaviour-Driven Development (BDD) has emerged in the last years as a powerful methodology to specify testable and executable user requirements through stories and scenarios. With the support of external testing frameworks, BDD stories can be used to automatically assess the behavior of a fully functional software system. This article describes a toolset which extends BDD with the aim of providing automated assessment also for user interface design artifacts to ensure their consistency with the user requirements since the beginning of a software project. The approach has been evaluated by exploiting previously specified user requirements for a web system to book business trips. Such requirements gave rise to a set of BDD stories that have been refined and used to automatically assess the consistency of task models, graphical user interface (GUI) prototypes, and final GUIs of the system. The results have shown that our approach was able to identify different types of inconsistencies in the set of analyzed artifacts and consistently keep the semantic traces between them.

## 1. Introduction

Modeling is recognized as a crucial activity to manage the abstraction and the inherent complexity of developing software systems. As a consequence, software systems tend to be designed based on several requirements artifacts which model different aspects and different points of view about the system. Considering that different phases of development require distinct information, resultant artifacts from modeling tend to be very diverse throughout the development, and ensuring their consistency is quite challenging [1]. To face this challenge, extra effort should be put on getting requirements described in a consistent way across the multiple artifacts. Requirements specifications should not, for example, describe a given requirement in a GUI prototype which is conflicting with its representation in a task model.

The key to successfully applying modeling in software development is to have effective communication between all project stakeholders. Because models are only abstract representations of software, abstractions that may not be accurate, development teams should strive to prove it with test code to show that your ideas actually work in practice and not just in theory. Developers, for example, rarely trust the software documentation, particularly detailed documentation, because it is usually out of sync with the code. That is the reason why executable specifications are always preferable over static specifications [2]. In practice, however, manual assessment and software inspections are usually the first approaches to assess software artifacts. Manually ensuring the consistency between requirements and artifacts every time a requirement is introduced and/or modified is, nevertheless, a discouraging activity for software development teams. Manual assessment is extremely time-consuming and highly error prone.

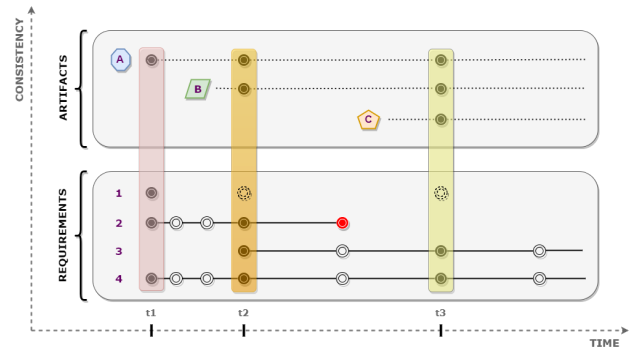

Figure 1. Snapshots of requirements and artifacts in different phases of the project.

An example of such a challenge is illustrated in Figure 1. Requirements and artifacts are supposed to evolve continuously along the project's lifecycle. When looking at different moments of a project's lifecycle, we should be able to guarantee that the current set of requirements at a given time is consistent with the current set of artifacts concerned by such requirements. For example, at time "t1", the project had 3 requirements (1, 2 and 4) and only 1 artifact (A) that should be consistent with them. At time "t2", requirements 2 and 4 have evolved and must be retested with respect to the artifacts. In addition, a new requirement (3) came up and a new artifact (B) was designed, so both artifacts A and B should be consistent with requirements 1, 2, 3, and 4. Later on, requirement 2 was dismissed and a new artifact (C) was introduced, so at time "t3", artifacts A, B, and C should be consistent with requirements 1 (not yet evolved), 3 and 4, but not anymore with requirement 2. That gives the dimension and the extent of challenges to consider when a large software system with multiple iterations is under development, not only because there are so many requirements and artifacts to align, but also because they come up, evolve, and are dismissed all along the project. Therefore, promoting automated assessment is a key factor to support verification and testing in an ever-changing environment.

Behaviour-Driven Development (BDD) [3] has aroused interest from both academic and industrial communities in the last years as a method allowing specifying testable user requirements in natural language using a single textual artifact [4], [5]. BDD describes User Stories [6] and their acceptance criteria in an easily understandable language for both technical and non-technical stakeholders. BDD stories allow specifying what we call "executable requirements", i.e. requirements that can be directly tested from their textual specification. Despite providing support to automated testing of user requirements by means of external testing frameworks, BDD and other testing approaches essentially focus on assessing fully functional software systems. Automated assessment of model-based artifacts such as task models, GUI prototypes, etc. is not supported.

Motivated by such a gap, we have researched and developed a tool-supported approach based on BDD to support the specification and the automated assessment of the functional aspects of user requirements on user interface design artifacts such as task models, GUI prototypes, and final GUIs. The contribution of this approach to both theory and practice in software verification and testing is to allow domain experts, requirements engineers, software testers, and other stakeholders to take advantage of an automated toolset to keep requirements and user interface design artifacts consistent throughout a software development process.

This article presents an extended and revised version of our approach previously published in [7]. The present article fully details the supporting toolset as well as the results we got in a case study exploiting user requirements previously specified as BDD stories by a group of potential Product Owners (POs). The stories have been used to automatically assess the user interface design artifacts for a web system to book business trips. The following sections present the foundations of this work (Section 2), an overview of our approach and its supporting toolset (Section 3), a case study we have conducted with the approach (Section 4), and an extended discussion of the results obtained with this case study (Section 5). We finish with our conclusion and perspectives for future works (Section 6).

## 2. Foundations and Related Work

### *2.1. Behaviour-Driven Development (BDD)*

According to Smart [8], BDD is a set of software engineering practices designed to help teams focus their efforts on identifying, understanding, and building valuable features that matter to businesses. BDD practitioners use conversations around concrete examples of system behavior to help understand how features will provide value to the business. BDD encourages business analysts, software developers, and testers to collaborate more closely by enabling them to express requirements in a more testable way, in a form that both the development team and the business stakeholders can easily understand. BDD tools can help turn these requirements into automated tests that help guide the developer, verify the feature, and document the application.

BDD specification is based on User Stories and scenarios which allow to specify executable requirements and test specifications by means of a Domain-Specific Language (DSL) provided by Gherkin[1]. User Stories were firstly proposed by Cohn [6]. North [9] has proposed a particular template to specify them in BDD and named it as "BDD story" (Figure 2).

```
Title (one line describing the story)

Narrative:
As a [role]
I want [feature]
So that [benefit]

Acceptance Criteria: (presented as Scenarios)

Scenario 1: Title
Given [context]
  And [some more context]...
 When [event]
 Then [outcome]
  And [another outcome]...

Scenario 2: ...
```

Figure 2. "BDD story" template.

In this template, BDD stories are described with a *title*, a *narrative* and a set of *scenarios* representing the acceptance criteria. The *title* provides a general description of the story, referring to a feature this story represents. The *narrative* describes the referred feature in terms of the role that will benefit from the feature ("*As a*"), the feature itself ("*I want*"), and the benefit it will bring to the business ("*So that*"). The acceptance criteria are defined through a set of *scenarios*, each one with a title and three main clauses: "*Given*" to provide the context in which the scenario will be actioned, "*When*" to describe events that will trigger the scenario, and "*Then*" to present outcomes that might be checked to verify the expected behavior of the system. Each one of these clauses can include an "*And*" statement to provide multiple contexts, events, and/or outcomes. Each statement in this representation is called a *step*.

BDD has been evaluated [10] and its characteristics analyzed and studied [11], [12] by several authors. Studies have been conducted to explore the use of BDD as part of empirical analysis of acceptance test-driven development [13], to support enterprise modeling within an agile approach [14] and within an user-centered approach [15], to support requirements engineering with gamification [16], to support a testing architecture for micro services [17], to support the analysis of requirements communication [18], to support safety analysis and verification in agile development [19], and to enhance the critical quality of security functional requirements [20]. Other studies have concentrated efforts in the use of automated acceptance testing to support BDD traceability [21], to provide traceability of BDD features to the source code [22], or to analyze its compatibility with business modeling [23], [24] and with BPMN [25].

---

[1] https://cucumber.io/docs/gherkin/reference/

BDD has also been used to support implementation of source code. Soeken et al. [26] propose a design flow where the designer enters into a dialog with the computer where a program processes, sentence by sentence, all the requirements creating code blocks such as classes, attributes, and operations in a BDD template. The template proposed by the computer can be revised; which leads to a training of the computer program and a better understanding of following sentences.

*2.2. User Interface Design Artifacts*

*2.2.1. Task Models*

Task models provide a goal-oriented description of interactive systems but avoiding the need of detail required for a full description of the user interface. Each task can be specified at various abstraction levels, describing an activity that has to be carried out to fulfill the user's goals. By modeling tasks, designers are able to describe activities in a fine granularity, for example, covering the temporal sequence of tasks to be carried out by the user or system, as well as any preconditions for each task [27]. The use of task models serves multiple purposes, such as better understanding the application under development, being a "record" of multidisciplinary discussions between multiple stakeholders, helping the design, the usability evaluation, the performance evaluation, and the user when performing the tasks. Task models are also useful as documentation of requirements both related with content and structure.

HAMSTERS [28] is a tool-supported graphical task modeling notation for representing human activities in a hierarchical and ordered manner. At the higher abstraction level, goals can be decomposed into sub-goals, which can in turn be decomposed into activities. The output of this decomposition is a graphical tree of nodes. Nodes can be tasks or temporal operators. Tasks can be of several types and contain information such as a name, information details, and criticality level. Figure 3 illustrates the task types in HAMSTERS. An abstract task is a task that involves sub-tasks of different types. A system task is a task performed only by the system. A user task is a generic task describing a user activity. It can be specialized as a motor task (e.g. a physical activity), a cognitive task (e.g. decision making, analysis), or perceptive task (e.g. perception of alert). Finally, an interactive task represents an interaction between the user and the system; it can be refined into input task when the users provide input to the system, output task when the system provides an output to the user, and input/output task that is a mix of both but performed in an atomic way.

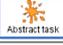

Figure 3. Task types in HAMSTERS.

Temporal relationships between tasks are represented by means of operators. The enable operator (>>) expresses that the tasks T1 and T2 occur sequentially, one after the other. Other operators are concurrent (|||), choice ([]), and order independent (|=|), that respectively express that tasks can be held simultaneously, the choice of one implies that the other will be disabled, and that the user can choose whether he will perform the one or another task first. The temporal operators allow extracting usage scenarios for the system. This is done by following the multiple achievable paths in the model, with each combination of them generating an executable scenario that can be performed in the system.

*2.2.2. Graphical User Interface (GUI) Prototypes and Final GUIs*

A GUI prototype is an early representation of an interactive system. They encourage communication, helping designers, engineers, managers, software developers, customers and users to discuss design options and interact with each other. Prototypes are often used in an iterative design process where they are refined and become more and more close to the final GUI through the identification of user needs and constraints. While the beginning of the project requires a low-level of formality with GUI prototypes being hand-sketched in order to explore design solutions and clarify user requirements, the development phase requires more refined versions, frequently describing presentation and dialog aspects of the interaction. By running simulations on prototypes, we can determine and evaluate potential scenarios that users can perform in the system [29].

Along this refining process, the prototype can be designed in different levels of fidelity. Based on that, the design of user interfaces is expected to evolve along the whole software development process. While the beginning of the project requires a low-level of formality with GUI prototypes being hand-sketched in order to explore design solutions and clarify user requirements, the development phase requires more refined versions frequently describing presentation and dialog aspects of interaction. Full-fledged versions of user interfaces are generally produced only later in the design process, and frequently corresponds to how the user "see" the system. From the user's point of view, the presentation aspect of a user interface actually is the system, so if some feature is not available there, then it does not exist. Mature GUI versions are the source for acceptance testing and will be used by users to assert whether or not features can be considered as done.

*2.3. Artifacts Assessment*

Since a long time ago, artifacts other than final versions of GUIs are usually only inspected manually in an attempt to verify its adequacy [30], [31]. Alternatively, requirements traceability techniques are employed as a way to trace such requirements along their representation in another artifacts [32]. Some approaches also concentrated efforts in providing automated tools to keep compatibility between requirements and their own artifacts. Luna et al. [33], for example, propose WebSpec which is a requirement artifact that can be used in conjunction with mockups to provide UI simulations, allowing some level of requirements validation, but not for GUI prototypes designed out of the approach. Buchmann and Karagiannis [34] presents a modeling method for the elicitation of requirements for mobile apps that enables semantic traceability for the requirements representation. The method, however, is not focused on GUI prototypes and can only validate requirements modeled within the approach. Solutions to generate GUIs from other software models, which in theory would keep them consistent, is also a topic that has received attention for long time [35].

There is also an intrinsic relationship between user interface design and task modeling, when considered in a user requirements perspective. Some authors have even tried to establish linguistic task modeling for designing user interfaces [36] where a notation enables identification of task input elements based on the task state diagram and dynamic tasks. Martinie et al. [37], followed by Campos et al. [38], also propose a tool-supported framework and a model-based testing approach to support linking task models to an existing, executable, and interactive application, defining a systematic correspondence between the GUI elements and the user tasks, but demanding a wide intervention in the source code of the application.

*2.4. Ontological Support*

The approach we propose in this article is supported by an ontology for describing interactive behaviors with a common vocabulary for writing BDD stories [39], [40]. The main benefit of this strategy is that BDD stories using this common vocabulary can support specification and execution of the scenarios directly on GUI-related artifacts. The ontology covers concepts related to presentation and behavior of interactive components used in web and mobile applications. It also models concepts describing the structure of BDD stories, tasks, scenarios, and GUIs.

This ontology has been designed based on several concepts borrowed from well-established works in the literature to describe interaction. Table 1 presents a comparison of the concepts used in the ontology with the ones used by other methods and languages. This analysis is presented for Cameleon Framework [41] and UsiXML [42], as well as for W3C MBUI Glossary [43] and SWC [44]. The Cameleon Reference Framework decomposes user interface design into a number of different components that seek to reduce the effort in targeting multiple contexts of use [41]. These components are Task-Oriented Specification, Abstract UI, Concrete UI and Final UI. The ontology has been built based on this decomposition, with high-level description of tasks being modeled as a task-oriented specification (based on notation such as CTT [45] and HAMSTERS [28]). UsiXML implements the Cameleon Framework in an XML specification, which allows us operating these concepts in the ontology. SWC adds the dialog component for the Cameleon/UsiXML specification allowing us specifying transitions and adding navigation to the UI. Finally, W3C MBUI Glossary contributes establishing the common vocabulary used by the other methods and languages. This common vocabulary is used to describe elements in the ontology. A full presentation of the ontology and its implementation is provided in [39], [40].

Table 1. A compared overview between the ontology and other methods and languages.

| | Concept | Mapping in the ontology |
|---|---|---|
| Cameleon and UsiXML | **Task-Oriented Specification**: This concept describes the tasks that the user and the system carry out to achieve the application's objectives. The tasks are described at a high level that is independent of how these are realized on a particular platform. | Description of **Scenario-based concepts**, including the modeling of **Users Stories** and **Tasks**. |
| | **Abstract UI**: This level describes models of the user interface that are independent of the choice of platform and the modes of interaction (visual, tactile, etc.). | Description of **Interaction Elements** in the **Presentation** perspective. |
| | **Concrete UI**: This level models the user interface for a given platform, e.g. desktop PC, tablet, smartphone, connected TV and so forth. | **Platform concepts** are described in the ontology, as well as the list of interaction elements that are supported by each platform (web and mobile). |
| | **Final UI**: This level implements the user interface for a specific class of device, e.g. an iPhone, or an Android tablet. | The ontology provides means of reading the set of **interaction elements** supported by each user interface platform. It allows designing automated testing implementations for specific platforms based on such elements to create concrete graphical widgets. |
| SWC | **Task Model (TM)**: Tasks and dependencies between tasks. | Description of **State Machine concepts**. The dynamic behavior of tasks being performed by users and systems are described as **Scenario-based concepts**. |
| | **Abstract User Interface (AUI)**: Relationship between logical presentation units (e.g. transition between windows), logical events, abstract actions. | Description of **Interaction Elements** in the **Dialog** perspective. |
| | **Concrete User Interface (CUI)**: States, (concrete) events, parameters, actions, controls, changes on UI dialog according to events, generic method calls, etc. | Description of the **Transition** triggers in the **State Machine** that each behavior may perform on the user interface. |
| | **Final User Interface (FUI)**: "Physical" signature of events, platform specific method calls, etc. | The ontology provides means of reading the set of **behaviors** supported by each interaction element. It allows designing automated testing implementations for specific platforms based on such behaviors to create concrete class methods for automating the "physical" interaction on the user interface. |
| W3C MBUI Glossary | It is a glossary of terms recurrent in the **Model-based User Interface domain (MBUI)**. It contains informal, commonly agreed definitions of relevant terms and explanatory resources. | Description and definition of **Platform** and **UI concepts**. |

## 3. A Tool-Supported Behavior-Based Approach

### 3.1. Assessment Strategy

An overview of our strategy for assessing the considered artifacts is illustrated through an example in Figure 4, where BDD scenarios are used to verify the consistency of the target artifacts (task models, GUI prototypes and final GUIs). Therein are exemplified five steps of scenarios being tested against equivalent tasks in task model scenarios, and the equivalent interaction elements on GUI prototypes and final GUIs. In the first example, the step *"When I select '<field>'"* corresponds to the task *"Select <field>"* in the task model scenario. Such a correspondence is due to the fact that the step and the task represent the same behavior, i.e. selecting something, and both of them are placed at the first position in their respective scenario artifacts. The interaction element "*field*" that will be affected by such a behavior will be assessed on the GUI prototype and on the final GUI. In both artifacts, such a field has been designed with a *CheckBox* as interaction element. The semantics of the interaction in *CheckBoxes* is compatible with selections, i.e. we are able to select *CheckBoxes*, so the consistency is assured.

The same is true in the example with the second step (*"When I click on '<field>'"*). There is a corresponding task *"Click on <field>"* at the same second position in the task model scenario, and the interaction element "*Button*", that has been chosen to address this behavior in both the GUI prototype and the final GUI, is semantically compatible with the action of clicking, thus the consistency is assured as well. In the third example, the step *"When I choose 'value' referring to 'field'"* is also compatible with the task *"Choose <field>"* in the task model, and with the interaction elements *DataChooser* and *Calendar*, respectively on the GUI prototype and on the final GUI. Notice that, despite being two different interaction elements, *DataChooser* and *Calendar* support a similar behavior, i.e. both of them support the behavior of choosing values referring to a field.

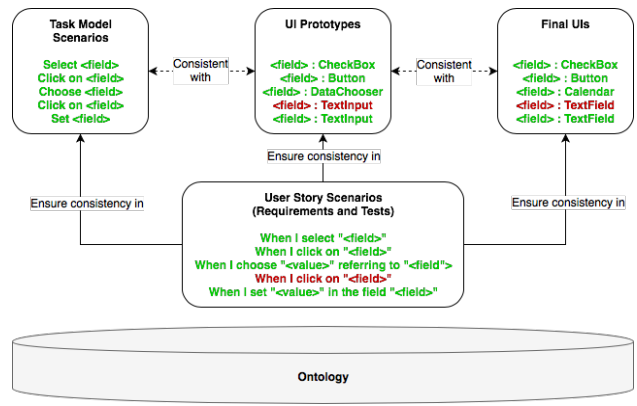

Figure 4. Illustrative example of the approach for assessing the different GUI artifacts.

The example provided with the fourth step (*"When I click on '<field>'"*) illustrates an inconsistency being identified. Even though there exists a corresponding task in the task model scenario, the interaction elements that have been chosen to address this behavior (*TextInput* in the GUI prototype and *TextField* in the final GUI) are not compatible with the action of clicking, i.e. such kind of interaction element does not semantically support such an action. The semantics of *TextInputs* (or *TextFields*) is receiving values, not being clicked. Such an example is provided by the fifth step (*"When I set 'value' in the field '<field>'"*). For this step, the consistency is assured because *TextInputs* and *TextFields* support the behavior of having values being set on them. All this semantic analysis is supported by the use of the aforementioned ontology that models the interaction elements and the interactive behaviors they support [39], [40]. The concept mapping table with the full list of supported interactive behaviors and the catalog of rules to assess each one of the artifacts is presented in the Appendix A at the end of this article.

The present strategy for assessment allows tracking some key elements in the user interface design artifacts and check whether they are consistent with the user requirements. When assessing these artifacts, we are mainly complying with the verification aspect of software testing, since by definition, we are comparing the requirements baseline with the successive refinements descending from it (i.e. the artifacts) in order to keep these refinements consistent with the requirements baseline. When assessing final GUIs, we are also complying with the validation aspect of software testing, once these artifacts are tested simulating the user's actions, thus checking if the software product satisfies or fits the intended use according to the user's acceptance criteria specified in the BDD stories.

This strategy has been implemented in Java EE and integrates multiple frameworks such as JBehave, JDOM, JUnit, and Selenium WebDriver. It is presented in detail hereafter along with the complete toolset supporting the approach.

### 3.2. Overview of the Toolset

The behavior-based approach we describe in this article is supported by a toolset that integrates a range of industry-accepted frameworks and libraries with the implementation of our strategy for assessing task models, GUI prototypes, and final GUIs. The general architecture of this toolset is presented in Figure 5 considering a particular setup of the strategy presented above. This setup considers BDD stories as the specific type of User Stories targeted, graphical user interfaces (GUIs) as the specific type of UIs, and web pages as the specific type of final UIs.

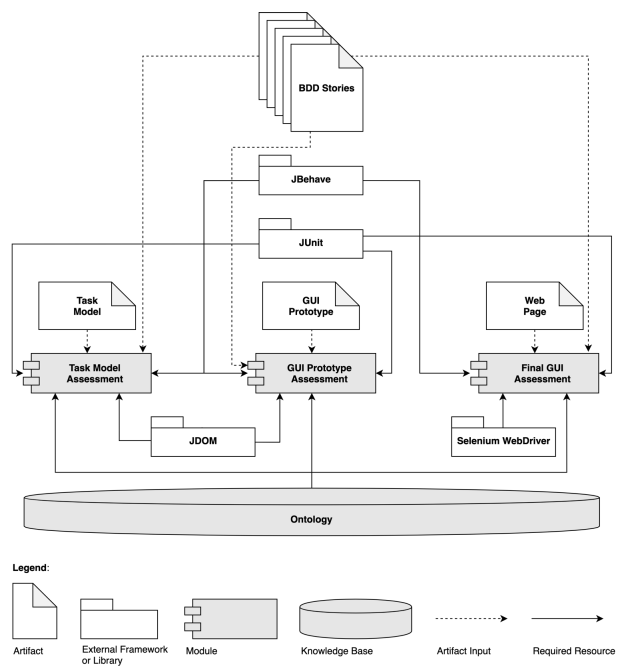

Figure 5. Toolset architecture proposed by our approach.

As the primary source of user requirements, BDD stories are the main input to the toolset. The stories are parsed by JBehave[2] which supports running stories as JUnit[3] tests. Task models, GUI prototypes, and web pages are the input artifacts to the respective assessment modules. JDOM[4] is used to parse the XML source files of task models and GUI prototypes, while Selenium WebDriver[5] is used to run the stories directly on a web browser. Finally, the aforementioned ontology is used as a knowledge base to provide the set of supported interactive behaviors and interaction elements to be assessed on the artifacts. The three assessment modules and how they integrate all these elements are described in detail hereafter.

### 3.3. Tool Support

#### 3.3.1. Task Model Assessment Module

Task models can be designed through a diverse set of notations and tools. We decided to implement a solution for HAMSTERS (v4.0) since the tool (*i*) allows the extraction of scenarios by running the model; and (*ii*) exports source files of both the reference model and the extracted scenarios in XML. The architecture we describe in this section is, however, flexible enough to accommodate other notation and tools in the future. For assessing the models, we parse the XML source file of the extracted scenarios produced by the HAMSTERS tool. The process consists of verifying, for each step in the BDD scenario, if there are one or more correspondences for such a step in the XML source files of the scenarios extracted from the task models. Checking this correspondence of steps in the BDD scenarios and tasks in scenarios extracted from task models is possible thanks to a formatting rule which assesses whether a behavior described in a step has an equivalent task to model it in the task model. This rule aims to eliminate unnecessary components of the step that do not need to be present in the task name. The mapping table presented in the Appendix A gives the full set of correspondences analyzed by our approach.

To do so, our testing algorithm fixes a step in the BDD scenario and retrieves from the ontology the corresponding task to be verified in the task model. Then we parse each task of each scenario in the XML source file looking for one or more correspondences to the task retrieved from the ontology. If matches are found, then a list of matches is created, keeping the position where each scenario-task match has been found. The algorithm presented in Figure 6 implements such a strategy.

```
foreach step from US Scenarios do
    taskToFind <- correspondent task from the ontology
    foreach task from each XML source file do
        if the attribute taskname is equal to taskToFind then
            ListOfMatches <- position(scenario,task)
        endif
    endforeach
endforeach
show ListOfMatches
```

Figure 6. Testing algorithm for assessing scenarios extracted from task models.

---

[2] https://jbehave.org

[3] https://junit.org

[4] http://www.jdom.org

[5] https://www.selenium.dev/

The results of testing are shown in a log indicating, for each step of the BDD scenario, if and where a given step has found an equivalent task in the XML file analyzed, and once it carries an object value associated, which value it is. As scenarios in BDD stories and scenarios in task models may be ordered differently, the algorithm checks the whole set of XML files to ensure we are looking for all the instances of the searched task. So, the log of execution shows, for each step of the BDD scenario, the results of searching in each XML scenario file (".scen"). Each line of results presents therefore:

- the name of the scenario in which the search has been carried out,
- the task name that has been searched for,
- the position at which the task has been found (if so), otherwise it is shown the message "Task not found!", and
- the object value associated with each task (if any), otherwise it is shown the message "No Value".

Due to that, if there are several XML files of scenarios, the results in the log will show where a correspondent task has been found in each one of them. A consequence of such a strategy is that the process of analyzing if a given task is correctly positioned in the evaluated scenarios is made manually after getting the whole log of results.

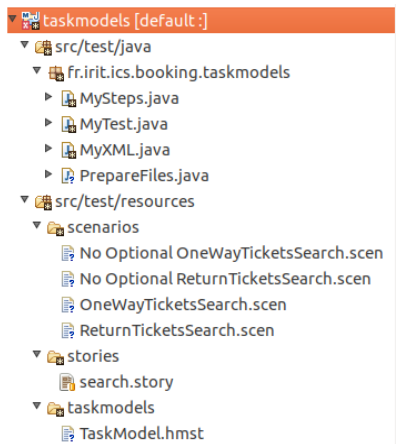

Figure 7. File tree for the implementation of task model assessment.

We have implemented this solution in the Eclipse IDE as a Java EE project using JDOM and JUnit. The project has been structured in two packages. The first one encompasses the classes for implementing the solution. As shown in Figure 7, this package contains four classes: *MySteps*, *MyTest*, *MyXML* and *PrepareFiles*. *MySteps* implements the mapping between the interactive behaviors described in the ontology and the assertion that should be made when checking scenarios from task models. *MyXML* implements methods for parsing scenario files extracted from task models in their XML files. *MyTest* is the JUnit class that triggers the set of BDD stories that have been selected for testing. Finally, *PrepareFiles* is the class in charge of preformatting the scenario source files extracted from task models. For this, before starting the assessment, we edit each scenario XML file to incorporate: (*i*) the name of the task referenced by each task ID; (*ii*) the information about the optionality of each referenced task; and (*iii*) the object value associated with each task, if it has been provided during the task execution. This is necessary because the HAMSTERS tool exports scenarios with only a reference to the task ID and the object ID that compose the flow. All the information is recovered from the reference task model XML file that actually contains the whole set of information about each task that has been modeled.

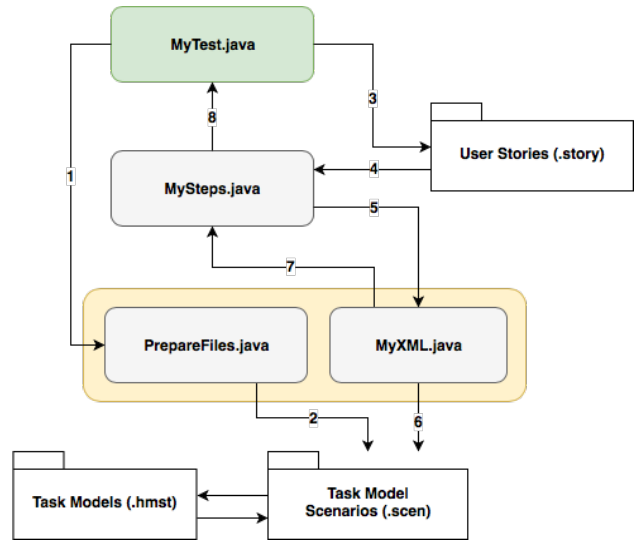

Figure 8. Flow of calls for running tests on task model scenarios.

The second package encompasses the resources demanded for running the tests. In the folder "*stories*", we have the whole set of BDD stories text files that have been specified for the project. Even being text files, each BDD story file must be named with a "*.story*" extension. In the example, the project has one single BDD story, with different scenarios for testing a given feature. The folder "*scenarios*" contains the current XML scenario files extracted from task models under testing, before and after the process of preformatting. Finally, the folder "*task models*" keeps the reference XML source files for the task models under testing. Such files are useful to allow the process of preformatting.

Figure 8 represents the flow of calls we have designed in our algorithm for running a battery of tests on task model scenarios. The flow starts with the class "*MyTest.java*". First of all, this class instantiates an object from "*PrepareFiles.java*" (flow 1) in order to trigger the process of preformatting mentioned before. Such a process runs on the package of task model scenarios (flow 2), naming the extracted tasks and adding useful complementary information for testing. For that, the process asks the reference source file (".*hmst*") of the corresponding task model mentioned by each task in the scenario. After getting the scenario files formatted, "*MyTest.java*" includes the BDD story (or the set of BDD stories) that will be tested (flow 3).

Each one of the steps in the BDD story under testing makes a call to the class "*MySteps.java*" (flow 4) that knows which behaviors are supported by the ontology. Based on the behavior referenced by the step, this class makes a call to the class "*MyXML.java*" (flow 5) in charge of parsing all the set of task model scenarios (flow 6). This parsing aims to check if the behavior addressed by the step is also present at the same position in at least one of the scenarios extracted from the task models. The result of this parsing is then returned to the class "*MySteps.java*" (flow 7). At this point, based on the algorithm presented in the previous section, a list of all the matches found during the parsing for each step is presented as a result. Finally, the class "*MySteps.java*" returns the result to the class "*MyTest.java*" (flow 8) that made the original call.

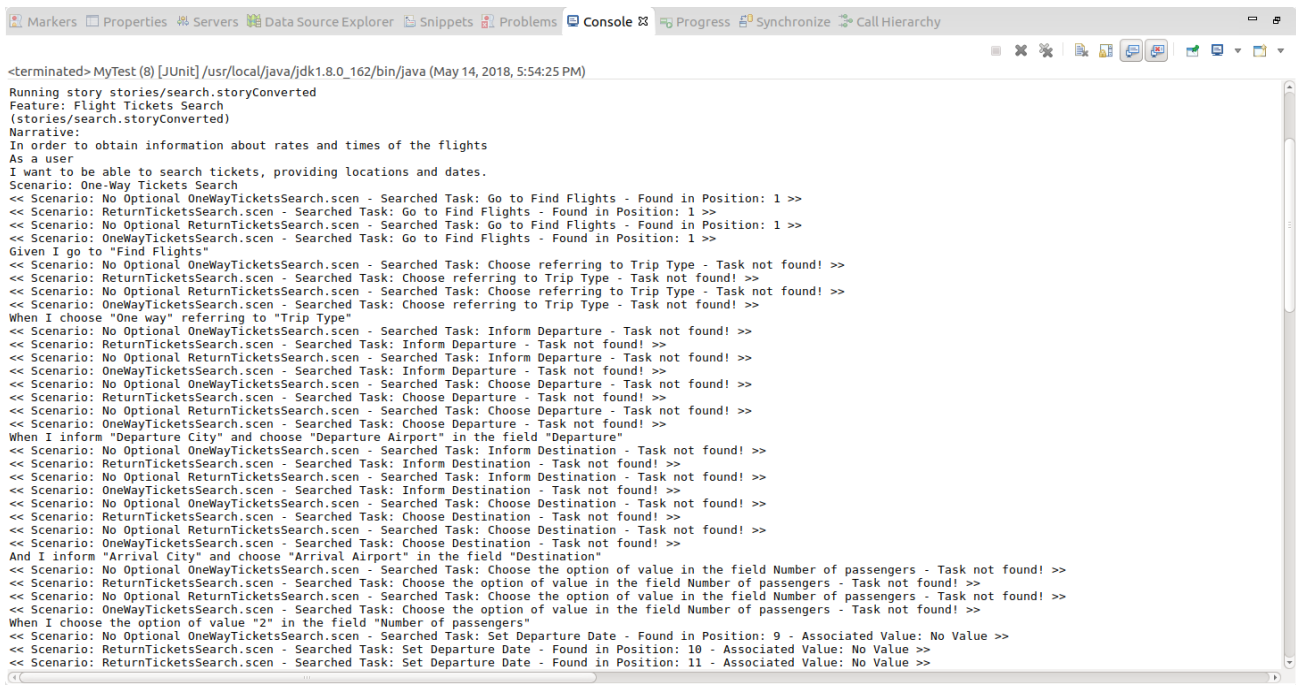

Figure 9. Console showing the log of results after running a BDD story.

Notice the independence of the components assigned at the core of the structure represented in Figure 8 (highlighted in yellow). Those components are related to the particularities of test implementation for HAMSTERS task models and scenarios. As mentioned before, "*PrepareFiles.java*" is in charge of preformatting the extracted scenario files and reading the reference source file of task models, while "*MyXML.java*" is in charge of parsing the scenario files, searching for the elements under testing. By doing this, we deliver a flexible architecture allowing, in the future, that task models and scenarios modeled by other modeling tools (or even by other versions of HAMSTERS) could also be tested by just implementing new interfaces for this core.

In summary, considering the presented architecture, to setup and run a battery of tests, one must:

- Place the set of task model scenario files (".*scen*") that will be tested in the package "*Task Model Scenarios*",
- Place the set of task model files (".*hmst*") that will support the test in the package "*Task Models*",
- Place the set of BDD story files (".*story*") that will be tested in the package "*User Stories*",
- Indicate in the "*MyTest*" class which BDD story will be tested, or which folder ("*/stories*") contains all the BDD stories that will be tested, and finally
- Run the "*MyTest*" class as a JUnit Test.

Thus, for running the tests, the "*MyTest*" class is triggered. This JUnit class specifies exactly which BDD story (or which set of BDD stories) will run (Figure 10). Finally, Figure 9 shows the console with the test results running the story shown in Figure 7. Notice that, for each step of the BDD scenario, it is shown where some corresponding task has been found and which value is associated with it (if any).

```
@Test
public void testAllStories() throws Throwable {
    eng.addSteps(new MySteps());
    eng.addStories("/stories/search.story");
    eng.run();
}
```

Figure 10. "*MyTest*" class indicating the file "*search.story*" for running.

### 3.3.2. GUI Prototype Assessment Module

Just like for task models, there are multiple notations and tools with different implementations for designing and modeling GUI prototypes [46], [47]. Among these multiple tools, we have chosen to implement a solution for Balsamiq in its version 2.2.28 once it fits our premise of producing visual wireframes ready for assessment. However, as we have done for the implementation of task models, we have designed a flexible and open architecture where other notations and tools could benefit from our approach by just implementing a new class in accordance with their own patterns to implement and model GUI prototypes.

The source code of Balsamiq prototypes is also provided by the use of an XML specification. Thus, our strategy for testing such prototypes is parsing their XML source files, looking for GUI elements that match the ontology description for each mapped behavior. The first step for assessing such prototypes is to get from the ontology the list of interaction elements that support the behavior mentioned by the BDD step under testing. After getting such a list of supported interaction elements, we pursue analyzing the Balsamiq XML file to identify, first of all, if the name of the interaction element present in the step appears in the prototype. This is made by reading the tag "<text>" identified in the parent tag "<controlProperties>" for a given "<control>" element. If such a field exists, i.e. there is a tag "<text>" carrying its name (case insensitive), so we retrieve which interaction element is associated with it. At this point, we have implemented a reference file containing the mapping between the abstract interaction elements in the ontology and the Balsamiq concrete implementation of such elements.

Balsamiq has two methods for representing interaction elements on its XML source files. They can be directly assigned with a unique "controlID" or be part of a group that encompasses a label and the interaction element itself. Our testing algorithm implements a solution that covers both situations. The algorithm presented in Figure 11 illustrates this implementation.

```
foreach step from US Scenarios do
    supportedUIElements <- correspondent UI Elements from the ontology
    fieldName <- name of the UI Element from the step
    foreach UI Element from the Balsamiq prototype do
        if the attribute text is equal to fieldName && is not in group then
            if the attribute controlTypeID is equal to one of the
               supportedUIElements then
                numElements++
            endif
        else if the attribute text is equal to fieldName && is in group then
            if the attribute controlTypeID of some member of the group is
               equal to one of the supportedUIElements then
                numElements++
            endif
        endif
    endforeach
endforeach
if numElements == 1 show Success
else show Fail
```

Figure 11. Testing algorithm for assessing Balsamiq GUI prototypes.

When looking for matching elements, the algorithm identifies which Balsamiq method has been used to design the element. If the parent tag is a label, it means that the element is part of a group that contains the element itself in a sibling tag. This sibling tag is then identified by reading the attribute "isInGroup". If the parent tag is not a label, so it is already the element itself. After identifying it, the algorithm checks if some of the interaction elements received from the ontology matches with the element from the prototype that is being investigated. If so, the variable "numTasks" is increased by one. After investigating the whole set of tags, the value of this variable is returned and must be equal to "1", which means only one interaction element for representing the "fieldname" has been found. If this value is equal to "0", it means that no interaction element has been found in the prototype with that "fieldname", while if it is greater than "1", it means that more than one interaction element has been found with the same "fieldname". In both cases, the algorithm identifies the failure and the test does not pass. This process is conducted for each step of the BDD scenario.

Notice that for prototypes at this level of refinement, we only assess the presentation aspect. We are not considering for testing at this level the dialog modeling and the consequent dynamic aspect of the interaction. It means that to check the consistency of the interaction elements modeled in the prototype, we only consider the presence (or the absence) of the right interaction elements on the screen where the interaction is supposed to occur. Behaviors that perform a state transition (e.g. navigating from one screen to another or getting mock values from the fields as a result of an interaction) are not being taken into account in the results.

Figure 12 represents the flow of calls we have designed in our algorithm for running a battery of tests on Balsamiq prototypes. The flow starts with the class "*MyTest.java*" that is a JUnit class in charge of triggering the battery of tests (its content is illustrated in Figure 13). This class indicates which files will be used for testing (flow 1). These files are distributed into two packages. The first one contains the BDD story files (where are the scenarios for testing), and the second one contains the Balsamiq GUI prototype files (that are the "*.bmml*" source files of Balsamiq prototypes). So, in the example of Figure 13, it has been indicated for testing the story "*Flight Ticket Search.story*" on the Balsamiq GUI prototype "*Book Flights.bmml*".

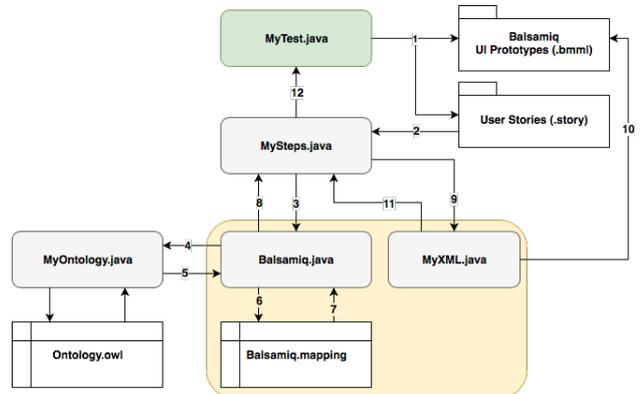

Figure 12. Flow of calls for running tests on Balsamiq prototypes.

```
@Test
public void testAllStories() throws Throwable {
    eng.addSteps(new MySteps("src/test/resources/lfprototypes/Book Flights.bmml"));
    eng.addStories("/stories/Flight Tickets Search.story");
    eng.run();
}
```

Figure 13. "*MyTest.java*": class for running tests on Balsamiq prototypes.

Each one of the steps in the BDD story under testing makes calls to the class "*MySteps.java*" (flow 2) that knows which behaviors are supported. Based on the behavior referenced by the step, this class makes a call to the class "*Balsamiq.java*" to get the list of Balsamiq interaction elements that supports such a behavior (flow 3). The class "*Balsamiq.java*" in its turn makes a call to the class "*MyOntology.java*" (flow 4) in charge of reading the OWL file of the ontology and recovering the list of abstract interaction elements supported by a given behavior. Such a list is then returned to the class "*Balsamiq.java*" (flow 5) that checks, for each element returned by the ontology, which are the corresponding concrete interaction elements from Balsamiq in charge of implementing the mentioned behavior (flow 6). This mapping is recovered from the file "*Balsamiq.mapping*" (flow 7).

Afterward, the class "*Balsamiq.java*" returns such a list with the concrete Balsamiq elements to the class "*MySteps.java*" (flow 8) that originally made the call. With the list of supported Balsamiq elements for the step under testing, the class "*MySteps.java*" makes a call to the class "*MyXML.java*" (flow 9) in charge of parsing the Balsamiq "*.bmml*" file (flow 10). This parsing aims to check if the prototype carries the interaction element mentioned in the step under testing, and if so, if such an element supports the behavior mentioned in the step. The result of this parsing is then returned to the class "*MySteps.java*" (flow 11). At this point, based on the algorithm presented in Figure 11, we verify how many instances have been found for the searched element. Finally, the class "*MySteps.java*" asserts the value and returns the result to the class "*MyTest.java*" (flow 12) that indicates if the test has failed or not.

Notice the independence of the components assigned at the core of the structure represented in Figure 12 (highlighted in yellow). Those components are related to the particularities of test implementation for Balsamiq prototypes. "*Balsamiq.java*" treats the demands for getting the corresponding abstract interactive elements from the ontology and translates them to the concrete interactive elements implemented by Balsamiq. "*Balsamiq.mapping*" provides such a translation. Finally, "*MyXML.java*" is in charge of parsing the "*.bmml*" Balsamiq files, searching for the element under testing. By doing this, we also deliver a flexible architecture allowing, in the future, that GUI prototypes modeled by other prototyping tools could also be tested by just implementing new interfaces for this core.

In summary, considering the presented architecture, to setup and run a battery of tests, one must:

- Place the set of "*.bmml*" files that will be tested in the package "*Balsamiq UI Prototypes*",
- Place the set of BDD story files ("*.story*") that will be tested in the package "*User Stories*",
- Indicate in the "*MyTest*" class which prototype will be tested with which BDD story (only a single prototype with a BDD story at a time), and finally
- Run the "*MyTest*" class as a JUnit Test.

### *3.3.3. Final GUI Assessment Module*

The testing of final GUIs we present in this section involves running tests directly on the web browser of a web application. As each environment has its own particularities, different testing frameworks would be required for performing tests on different environments. Despite our ontology supports a specification for both web and mobile environments, so far, we have only implemented an architecture to perform tests on a web environment, i.e. running on a web browser.

Besides using a framework to control navigation on a web browser, other frameworks are required to parse the sentences of the BDD stories, to build the test suit, and to generate reports from the execution. To test final GUIs directly from BDD stories, we use external frameworks to provide automated execution on the final GUI. Such frameworks are able to mimic user interactions with the final GUI by running the set of scenarios described in the BDD stories. Therefore, we have built an architecture of tools to put together the multiple set of required frameworks for performing our testing approach on web final GUIs. Such an architecture is presented hereafter.

The integrated tools architecture we propose is essentially based on Demoiselle and JBehave, Selenium WebDriver, and JUnit. We use Selenium WebDriver to run navigational behavior, and Demoiselle and JBehave to parse the scenario script. Test results provided by the JUnit API indicate visually which tests passed and which ones failed and why. Execution reports of BDD stories, scenarios and steps can also be obtained by using the JBehave API.

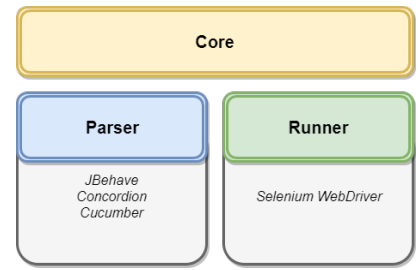

Figure 14. A 3-module integrated tools architecture.

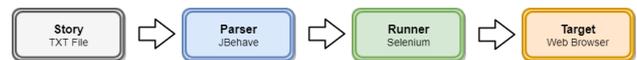

Figure 15. Flow of components in the proposed architecture.

This architecture allows developers and testers to automate testing on web user interfaces by following our behavior-based approach. The architecture has three main modules: *Core*, *Parser*, and *Runner* (Figure 14). *Core* is responsible for the main interfaces of the framework by orchestrating the information among the other 3 modules. The *Parser* is responsible for the abstraction of the component that will transform the story into Java code, to send to the *Runner* through standard or project-specific sentences. The *Runner* is responsible for the abstraction of the component that will perform navigation on the GUI, such as Selenium WebDriver or directly JUnit. The framework identifies stories written in "*.txt*" files to be sent to the *Parser* module and later to the *Runner*, which is responsible for interacting with a web browser using the Selenium WebDriver. This flow is illustrated in Figure 15. To run tests in such an architecture, story files are charged as inputs for the parser that translates the natural language behaviors into Java methods and then selects a runner to perform the navigational commands on a given targeted web browser.

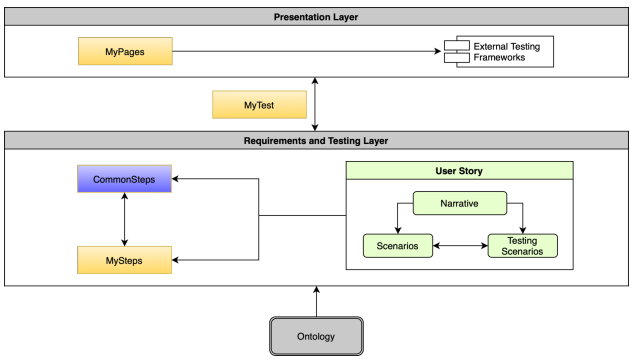

Figure 16. Packages and classes being structured to implement our testing approach.

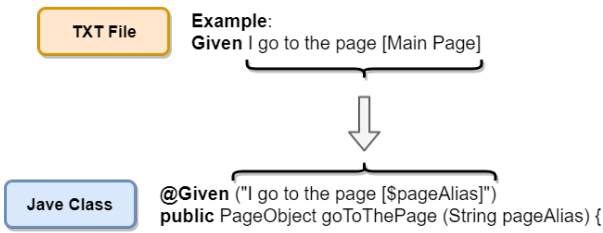

Figure 17. Parsing a step from a ".*txt*" BDD step to a Java method.

Figure 16 details how we have structured packages and classes in different layers to implement our architectural approach. The ontology provides to the model a pre-defined set of behaviors used in the "*Requirements and Testing Layer*". Interactive behaviors charged from the ontology are implemented by the "*CommonSteps*" class. New extended behaviors, that are not initially covered by the ontology, can be implemented in the "*MySteps*" class. Steps in BDD stories are mapped to either "*CommonSteps*" or "*MySteps*" behaviors in order to be run as Java methods. Figure 17 illustrates this mechanism.

The "*Presentation Layer*" includes the "*MyPages*" class which implements the link between the abstract GUI components defined in the ontology and the concrete GUI components instantiated on the interface under testing. This link is crucial to allow the Selenium WebDriver and other external testing frameworks to automatically run scenarios in the right components on the GUI. To link these components, the "*MyPages*" class identifies a screen map ("*@ScreenMap*") which address the web page location, and several element maps ("*@ElementMap*") which link the various abstract GUI elements in the BDD stories to their concrete GUI siblings on the user interface. This link is made by manually associating the name of each abstract GUI element with their concrete locators (such as IDs, XPaths, or any other web element identifier). Figure 18 illustrates this mechanism.

Finally, the "*MyTest*" class is a JUnit class in charge of triggering the tests, pointing which scenarios should be executed at a time, besides making the bridge between GUI components in in the Presentation Layer and executable behaviors in the Requirements and Testing Layer.

These three basic classes ("*MySteps*", "*MyPages*", and "*MyTest*") can also be modeled within the packages "*steps*", "*pages*" and "*tests*", in order to separate concerns and implement different classes for different pages or features. The structure of the Java project is presented on the left side of Figure 19. Notice that the three aforementioned classes are packed in the package "*java*" and the BDD stories in the package "*resources*". On the right side of the figure, the structure of the "*MyTest*" class is presented, highlighting the addition of new extended behaviors in the "*MySteps*" class as well as the stories (in the "*/stories*" folder) which will be triggered by a JUnit test method.

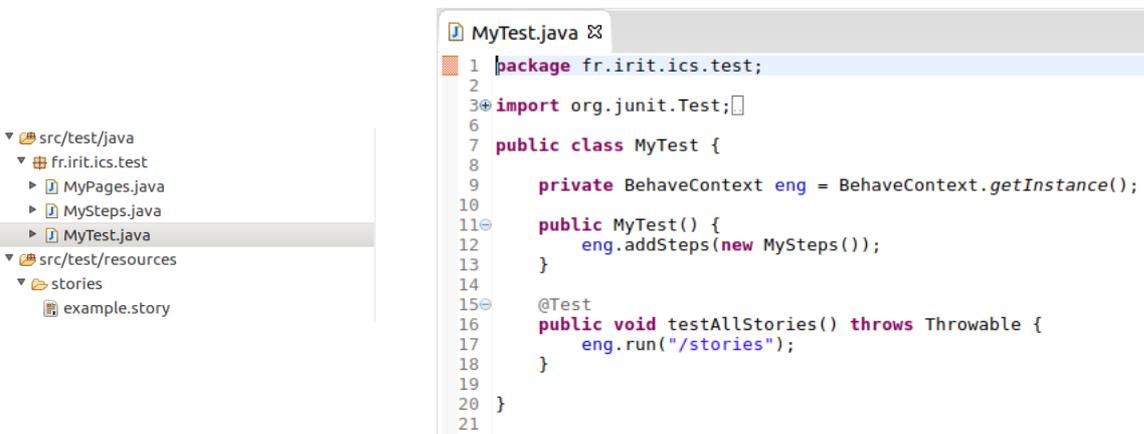

Figure 18. "*MyPage*" Java class.

Figure 19. Package tree (on the left) and "*MyTest*" class (on the right).

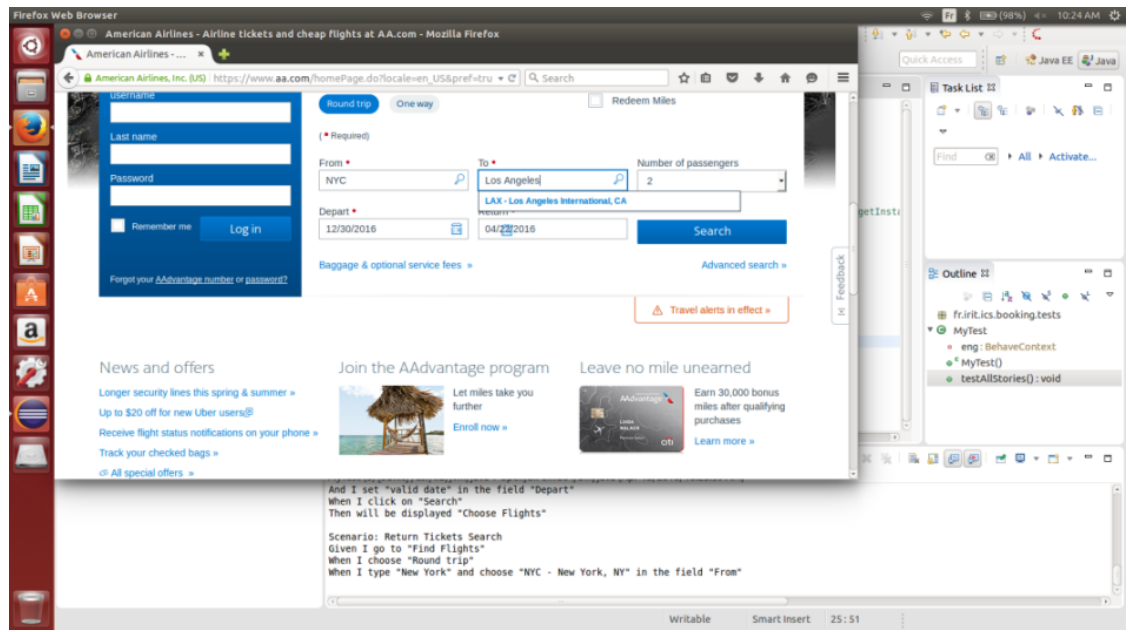

Figure 20. Automated execution of a scenario for searching airline tickets.

The environment for implementing and running the tests is the Eclipse IDE with a Maven project instantiated. Figure 20 shows such an environment with the "*MyTest*" class automatically running a scenario for searching airline tickets on the American Airlines website.

A resource that facilitates the writing of BDD stories (and that is also available when writing and editing BDD stories for assessing task models and GUI prototypes) is the immediate feedback concerning the existence of behaviors in the ontology to address the step that is being written. Figure 21 illustrates this resource. Notice that all the steps in the scenario have been recognized, i.e. there are equivalent behaviors in the ontology to address them, except the step "*When I set the date '12/20/2017' in the field 'Return'*" that has been underlined to alert that such a step is not recognized by the ontology (actually the correct step in this case is a generic: "*… I set '<value>' in the field '<element>'*" as used in the following line). When clicking on the alert icon, a message to say that "no step is matching" will be shown. Additional feedback is also given recognizing in the step the mention to values and interactive elements when they are surrounded by quotation marks.

The testing results are presented through the classical JUnit green/red bar within the Eclipse IDE. By the end of the tests, a JBehave detailed report is automatically generated in the project folder. Additionally, for each error found, screenshots are taken and stored to allow a better analysis of the results afterwards. Examples of these features are presented in Figure 22.

### 3.4. Alternatives for Performing the Approach

Depending on which phase the project is, our approach can be applied in two ways. Figure 23 illustrates these alternatives. The first case (represented on the left side of the figure) concerns the one where our approach will be implemented when the project is already running, and artifacts have already been designed (2). In such a case, our approach can be used to assess such artifacts, indicating where they are not in accordance with the specified requirements. In this case, requirements are supposed to be already identified (1), so we can directly write the BDD stories from these requirements (4), and likewise extract scenarios from the scenarized artifacts (3) (in our approach, the user interface design artifacts considered). When doing that, tests will be ready for running (5).

The second case (represented on the right side of the figure) refers to a project in the beginning, where no artifacts have been designed yet. In this case, by following the interactive behaviors covered by the ontology, the artifacts can be modeled in a consistent way from the beginning, already taking into account the possible interactions supported by each interaction element on the GUI. For this case, we could follow sequential steps that include: (1) identify the requirements, (2) design the scenarized artifacts from these requirements, (3) extract scenarios from these artifacts, (4) write formatted BDD stories based on the extracted scenarios, and finally (5) run tests on the artifacts. Alternatively, we can perform activity 4 (write formatted BDD stories) before activity 2 (design scenarized artifacts). It means that depending on the characteristics of the project, either the BDD stories can support the design of the artifacts, or the artifacts (by means of their extracted scenarios) can support the writing of BDD stories.

Figure 24 illustrates the resultant graph of options considered. The colored lines indicate the possible paths to be taken in the workflow. The yellow path indicates the design of scenarized artifacts before writing formatted BDD stories. The green path indicates the opposite, while the blue path indicates both activities in parallel. Notice

```
30 Scenario: Return Tickets Search
31 Given I go to "Find Flights"
32 When I choose "Round trip"
33 And I type "New York" and choose "NYC - New York, NY" in the field "From"
34 When I type "Los Angeles" and choose "LAX - Los Angeles International, CA" in the field "To"
35 And I choose the option of value "1" in the field "Number of passengers"
36 When I set the date "12/15/2017" in the field "Depart"
37 And I set "12/20/2017" in the field "Return"
38 When I click on "Search"
39 Then will be displayed "..."
```

Figure 21. Writing a BDD story and getting instant feedback of unknown steps.

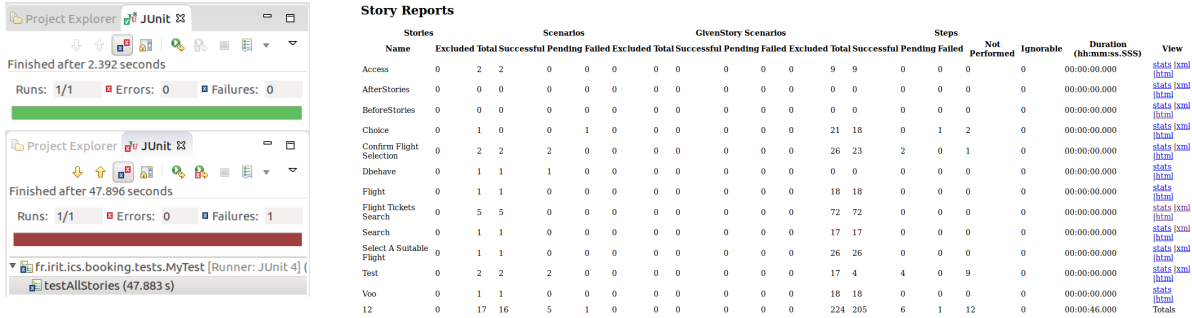

Figure 22. JUnit green/red bar on the left, and JBehave detailed report on the right.

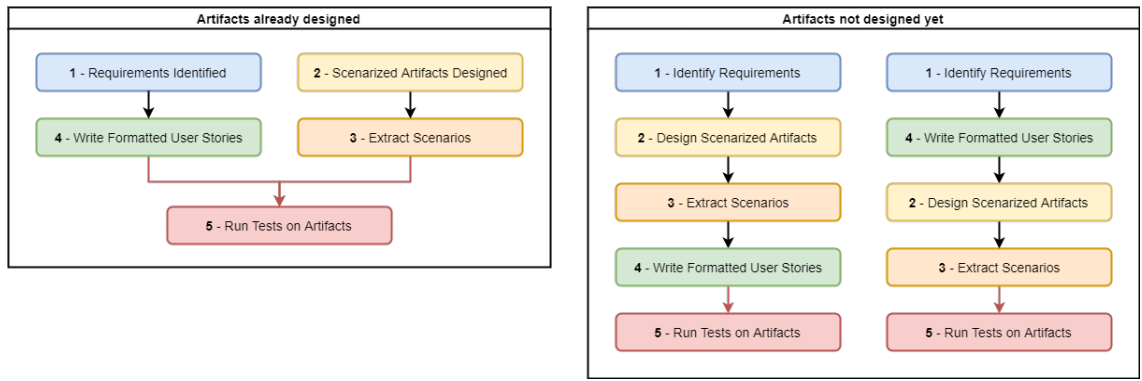

Figure 23. Alternatives for performing the approach (colors are used only to visually identify the same activities).

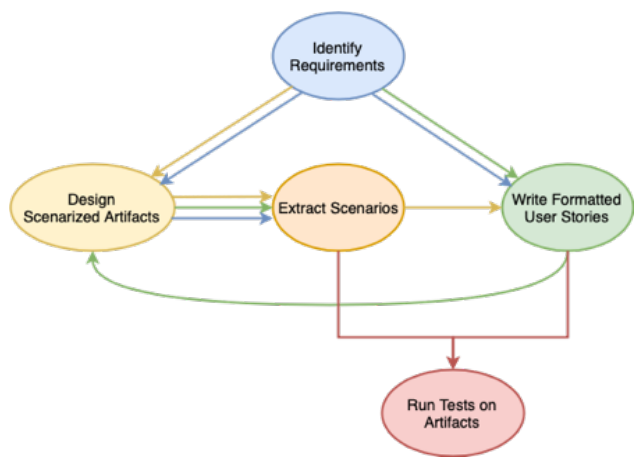

Figure 24. The graph of options for performing our approach (colors are used to visually identify the different paths).

that regardless of the path chosen, the extraction of scenarios is only possible after having designed the scenarized artifacts, and the identification of requirements is a precondition for all the other activities. Finally, to run tests on the artifacts, it is required to have extracted scenarios and written the BDD stories. The approach also benefits from the independence for testing artifacts, i.e., tests can run on a single artifact or on a set of scenarized artifacts which will be targeted at a given time.

## 4. Case Study

To investigate the potential of the approach, we conducted a case study with an existing web system for booking business trips. We studied the current implementation of user requirements in this system, and by applying a manual reverse engineering, we redesigned the appropriate task models and GUI prototypes for that system. We then exploited a set of BDD stories collected in a previous study [48] to simulate the assessment of the resultant user interface design artifacts. The aim of the present study is to provide an evaluation regarding the extent of inconsistencies our approach can identify in the targeted artifacts.

## 4.1. Methodology

We started the study by setting up an initial version of the BDD stories and their test scenarios to act respectively as our user requirements and acceptance criteria. We then reverse reengineered initial versions of task models and GUI prototypes from the existing web system. After getting ready a first version of task models, we extracted a representative set of scenarios from them. By following our strategy for testing, we parsed and ran this initial version of the BDD stories against the initial set of scenarios extracted from task models. Results were then evaluated, and we could observe the inconsistencies that were identified. As the strategy we follow for assessing scenarios in both task models and BDD stories parses all the steps of each scenario at once, the first round of results was obtained with a single battery of tests.

Following this step, we ran the same initial version of the BDD stories against initial versions of GUI prototypes designed using Balsamiq. Unlike the strategy for assessing task models, the strategy we follow for assessing GUI prototypes and final GUIs parses each step of each scenario at a time, so if an error is found out, the test stops until the error is fixed. That requires to run several batteries of tests until having the entire set of scenarios tested. Consequently, at the end of running, the tested scenarios are fully consistent with the respective GUIs.

We applied the same strategy to test the final GUIs, once they are fully-fledged versions of previously developed GUI prototypes. The difference here is that, by applying a reverse engineering approach, we assume that the released version of the current booking system (and of course its final GUIs) represents the unequivocal statement of the user requirements, once for the purpose of this study, we cannot modify them. As such, we did not have the opportunity to eventually redesign the final GUI to comply with the BDD stories we had set up. As a consequence, all identified inconsistencies necessarily resulted in modifications in the BDD steps, not in the final GUI.

Finally, we analyzed the results of testing in each artifact by mapping such results to identify the trace of each inconsistency throughout the artifacts. That gave us a complete traceability overview of each step of the BDD scenarios and their representation on the targeted artifacts.

## 4.2. Results

In total, we set up for assessment 3 BDD stories with 15 different scenarios, reengineered 3 task models (and extracted 10 scenarios from them), reengineered 11 GUI prototypes, and tested 7 different final GUIs. When presenting the results below, for the scenarios extracted from task models, the expected results are the positions of each correct equivalent task in such scenarios, while the actual results are the position at which the task was actually found. For GUI prototypes, the expected result for each step always equals to 1, i.e. there is one and only one interactive element in the prototype to address a given behavior from the step. The actual results are the number of times a semantically equivalent element has been found in the prototype to address such a behavior. Finally, for final GUIs, the expected results are the elements on the GUI which are expected to address a given behavior from the step, whilst the actual results are the actual elements found or the reason of test failure.

Table 2 presents, for each considered artifact, the mapping of results for the first battery of tests in each step of the full scenario for booking a roundtrip. As some steps were being updated after having previously failed in a given artifact, the results shown in yellow on the table indicate that, for the artifact in question, the test run with an updated version of the step and still failed. Results shown in blue indicate that, for the artifact in question, the test run with an updated version of the step and the test passed with such a version. Results shown in green indicate steps that passed the test in that artifact, and results in red indicate steps that failed in that artifact. Finally, results shown in orange indicate that the execution of such a step has been pending in that artifact, it is the case of the steps that effectively conclude the booking on the final GUI. We avoided such steps to not create fake reservations in the targeted booking system which was in production. In the column BDD stories/scenarios, we considered the original steps, as conceived before starting the first battery of tests in any artifact. Notice that once some step of scenario for some artifact fails, the scenario is considered as failed as well. Figure 25 additionally shows, as an example, a schematic representation of some BDD steps being assessed on specific elements of the considered artifacts.

Analyzing the results of mapping presented above for the first scenario "Successful Roundtrip Tickets Search With Full Options", we notice that the first step (that has succeeded in the task model) failed when tested with the Balsamiq prototypes. The reason is that the prototype had not addressed the web pages correctly, i.e. the "Book Flights" page could not be identified there. In a following battery of tests, this page has ended up being named "Flight Search" instead, which made the test passes when running on the final GUI.

Table 2. Mapping of the results after testing.

| BDD stories | Scenario from Task Model | Balsamiq GUI Prototype | Final GUI |
|---|---|---|---|
| **User Story**: Flight Tickets Search<br><br>**Narrative**:<br>**As an** IRIT researcher<br>**I want to** be able to search air tickets for my business trips, providing destinations and dates<br>**So that** I can obtain information about rates and times of the flights. | - | - | - |

| BDD stories | Scenario from Task Model | Balsamiq GUI Prototype | Final GUI |
|---|---|---|---|
| **Scenario: Successful Roundtrip Tickets Search With Full Options** | **FAILED** | **FAILED** | **FAILED** |
| **Given** I go to "Book Flights" | **Expected**: 1 | **Expected**: 1 | **Expected**: Flight Search |
| | **Actual**: 1 | **Actual**: 0 | **Actual**: Flight Search |
| **When** I inform "Toulouse" and choose "Toulouse, Blagnac (TLS)" in the field "Departure" | **Expected**: 2/3 | **Expected**: 1 | **Expected**: Departure |
| | **Actual**: 0 | **Actual**: 1 | **Actual**: Departure |
| **And** I inform "Paris" and choose "Paris, Charles-de-Gaulle (CDG)" in the field "Destination" | **Expected**: 4/5 | **Expected**: 1 | **Expected**: Destination |
| | **Actual**: 0 | **Actual**: 1 | **Actual**: Destination |
| **When** I set "Sam, Déc 1, 2018" in the field "Departure Date" | **Expected**: 6 | **Expected**: 1 | **Expected**: Departure Date |
| | **Actual**: 8 | **Actual**: 0 | **Actual**: Departure Date |
| **And** I set "08:00" in the field "Departure Time Frame" | **Expected**: 7 | **Expected**: 1 | **Expected**: Departure Time Frame |
| | **Actual**: 9 | **Actual**: 1 | **Actual**: Element not identified |
| **When** I choose "Round Trip" | **Expected**: 8 | **Expected**: 1 | **Expected**: Round Trip |
| | **Actual**: 0 | **Actual**: 1 | **Actual**: Round Trip |
| **And** I set "Lun, Déc 10, 2018" in the field "Arrival Date" | **Expected**: 9 | **Expected**: 1 | **Expected**: Arrival Date |
| | **Actual**: 10 | **Actual**: 0 | **Actual**: Arrival Date |
| **When** I set "10:00" in the field "Arrival Time Frame" | **Expected**: 10 | **Expected**: 1 | **Expected**: Arrival Time Frame |
| | **Actual**: 11 | **Actual**: 1 | **Actual**: Element not identified |
| **And** I choose the option of value "2" in the field "Number of Passengers" | **Expected**: 11 | **Expected**: 1 | **Expected**: Number of Passengers |
| | **Actual**: 12 | **Actual**: 0 | **Actual**: Element not found in "Flight Search" |
| **When** I set "6" in the field "Timeframe" | **Expected**: 12 | **Expected**: 1 | **Expected**: Timeframe |
| | **Actual**: 0 | **Actual**: 0 | **Actual**: Element not identified |
| **And** I select "Direct Flights Only" | **Expected**: 13 | **Expected**: 1 | **Expected**: Direct Flights Only |
| | **Actual**: 14 | **Actual**: 0 | **Actual**: Direct Flights Only |
| **When** I choose the option of value "Economique" in the field "Flight Class" | **Expected**: 14 | **Expected**: 1 | **Expected**: Flight Class |
| | **Actual**: 0 | **Actual**: 0 | **Actual**: Element not identified |
| **And** I set "Air France" in the field "Companies" | **Expected**: 15 | **Expected**: 1 | **Expected**: Company 1 |
| | **Actual**: 0 | **Actual**: 0 | **Actual**: Value does not fit the field |
| **When** I submit "Search" | **Expected**: 16 | **Expected**: 1 | **Expected**: Search |
| | **Actual**: 17 | **Actual**: 1 | **Actual**: Search |
| **Then** will be displayed "2. Sélectionner un voyage" | **Expected**: 17 | **Expected**: 1 | **Expected**: Proper Message |

| BDD stories | Scenario from Task Model | Balsamiq GUI Prototype | Final GUI |
|---|---|---|---|
| | **Actual**: 0 | **Actual**: 0 | **Actual**: Proper Message |
| **User Story**: Select a Suitable Flight<br><br>**Narrative**:<br>**As an** IRIT researcher<br>**I want** to get a list of compatible flights (including their rates and times) in accordance with my search criteria<br>**So that** I can select a suitable flight based on my needs. | - | - | - |
| **Scenario: Select a Return Flight Searched With Full Options** | **FAILED** | **FAILED** | **PASSED** |
| **Given** "Availability Page" is displayed | **Expected**: 18 | **Expected**: 1 | **Expected**: Availability Page |
| | **Actual**: 0 | **Actual**: 0 | **Actual**: Availability Page |
| **When** I click on "No Bag" referring to "Air France 7519" | **Expected**: 19 | **Expected**: 1 | **Expected**: Air France 7519 |
| | **Actual**: 0 | **Actual**: 0 | **Actual**: Air France 7519 |
| **And** I click on "No Bag" referring to "Air France 7522" | **Expected**: 20 | **Expected**: 1 | **Expected**: Air France 7522 |
| | **Actual**: 0 | **Actual**: 0 | **Actual**: Air France 7522 |
| **When** I click on "Book" | **Expected**: 21 | **Expected**: 1 | **Expected**: Book |
| | **Actual**: 0 | **Actual**: 1 | **Actual**: Book |
| **Then** will be displayed "J'accepte les Conditions d'achat concernant le(s) tarif(s) aérien(s)." | **Expected**: 22 | **Expected**: 1 | **Expected**: Proper Message |
| | **Actual**: 0 | **Actual**: 0 | **Actual**: Proper Message |
| **User Story**: Confirm Flight Selection<br><br>**Narrative**:<br>**As an** IRIT researcher<br>**I want** to get all the required data to confirm my flights<br>**So that** I can check the information, the fare rules and then finalize my booking. | - | - | - |
| **Scenario: Confirm a Flight Selection (Full Version)** | **FAILED** | **FAILED** | **PENDING** |
| **Given** "Confirmation Page" is displayed | **Expected**: 23 | **Expected**: 1 | **Expected**: Confirmation Page |
| | **Actual**: 0 | **Actual**: 0 | **Actual**: Confirmation Page |
| **When** I choose "I accept the General Terms and Conditions." | **Expected**: 24 | **Expected**: 1 | **Expected**: Proper Field |
| | **Actual**: 0 | **Actual**: 1 | **Actual**: Proper Field |
| **And** I click on "Finalize the trip" | **Expected**: 25 | **Expected**: 1 | **Expected**: Finalize the trip |
| | **Actual**: 0 | **Actual**: 0 | **Actual**: Finalize the trip |
| **Then** will be displayed "Votre voyage a été confirmé!" | **Expected**: 26 | **Expected**: 1 | **Expected**: Proper Message |
| | **Actual**: 0 | **Actual**: 0 | **Actual**: Proper Message |

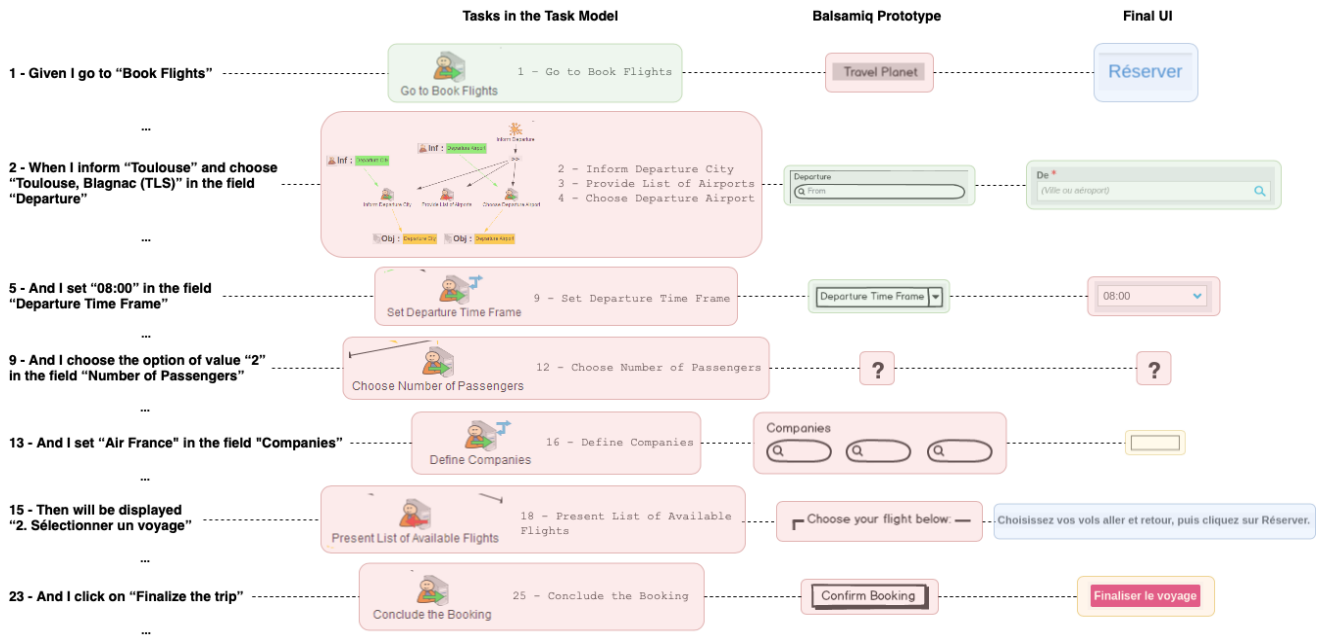

Figure 25. Schematic representation of User Stories steps being assessed on elements of the considered artifacts.

The two following steps have failed for task models but passed for Balsamiq prototypes and the final GUI. The reason of failure in task models is due to the additional tasks "*Provide List of Airports*" for the group of tasks which provides information about departures and destinations in the task model, from the second step until the eighth step, the gap between the expected position and the actual one was exactly two positions. However, both steps passed when tested for the Balsamiq prototype and the final GUIs, once the GUI element was correctly represented, i.e. as a "*SearchBox*" in the prototype, and as an "*Autocomplete*" field in the final GUI. The step testing the field "*Departure Date*" nonetheless failed for the Balsamiq prototype but passed for the final GUI (the same has occurred with the field "*Arrival Date*"). The reason of failure in the prototype is that the label of the field and the GUI element itself were not represented in the same group of elements. Contrasting with that, the step testing the field "*Departure Time Frame*" passed for the Balsamiq prototype but failed for the final GUI (the same has occurred with the field "*Arrival Time Frame*"). The reason is that it was not possible to locate unique identifiers for the element on the final GUI.

At the sixth step (*"When I choose 'Round Trip'"*), as the task for choosing the "*Round Trip*" was not exported from the task model to the scenario, the gap from the eighth position in task model scenarios until the end of the scenario (excluding the tasks not found) dropped down from two to one position. Such step succeeded when identifying the element "*Round Trip*" in the Balsamiq prototype and in the final GUI. The element "*Number of Passengers*" at the ninth step was not found both in the Balsamiq prototype and in the final GUI, despite being specified in the task model scenario (although not in the right position). The element "*Timeframe*" at the tenth step was written as "*Time Frame*" in the Balsamiq prototype and was misspecified with an unknown behavior "*Adjust Timeframe*" in the task model, so it could not be identified in these artifacts. The same has occurred with the element "*Flight Class*" at the twelfth step which was written as just "*Class*" in the Balsamiq prototype and specified as "*Define Fight Class*" in the task model, an unknown behavior. In the final GUI, they have not been identified as well, but due to the problem of unique identifiers.

The element "*Direct Flights Only*" at the following step was written as "*Only direct flights*" in the Balsamiq prototype, so it was not identified. It was correctly written in the final GUI and then it was correctly identified there. The corresponding task for this step was at the wrong position in the task model scenario. The element "*Companies*" was misspecified with an unknown behavior "*Define Companies*" in the task model. In the Balsamiq prototype, it was addressing three different "*SearchBoxes*", so it could not be identified as a unique and single element. After both the corresponding step and the prototype are redesigned to identify each field separately, the step should pass the test in the final GUI, but it ended up failing as well because the value informed on it ("*Air France*") did not fit the correspondent *Text Field* which only accepted 3 characters. In this case, the step was fixed to inform only the value "*AF*", the corresponding code defined to be used on the final GUI.

The button "*Search*" was correctly identified both in the Balsamiq prototype and the final UI, but the referenced task in the scenario extracted from the task model was found at the wrong position. Finally, for the first scenario, the message resultant from the user interaction when searching flights was not identified in the scenario tasks and was not reachable in the Balsamiq prototype due to the untraceable interaction between screens. In the final GUI, the message to check was modified, so the step was refactored to reference the new message. Thereby, after the modification, the test passed for this artifact.

For the steps in the second and third scenarios, all of them failed for the task model and a deep work for fixing the compatibility issues would be required. Specification of tasks did not follow the behaviors mapped in the ontology, so none of them could be identified during the test. The prototypes of web pages that should be displayed when starting those scenarios failed once they addressed wrong page names. They were correct in the final GUIs, nonetheless. The choice of flights in the second and third steps of the second scenario failed in the Balsamiq prototype (the name of the element was misidentified) and were refactored to require the action of clicking on the number of the flights (instead of on the fare profile), so they passed the test in the final GUI. The behavior of clicking on the button "*Book*" was correctly addressed in both the Balsamiq prototype and in the final GUI. The verification of message after the interaction was succeeded in the final GUI but failed as expected in the Balsamiq prototype due to the untraceable interaction between screens.

In the third and last scenario, the behavior of accepting the general terms and conditions to confirm the booking of flights was correctly addressed in both the Balsamiq prototype and the final GUI. The button "*Finalize the trip*" was not identified in the Balsamiq prototypes (it was "*Confirm Booking*"). As stated before, the action of clicking the button and verifying the confirmation message in the final GUI was not performed in order to not effectively place a fake booking in the system. Due to that, both steps were signalized as "pending" in the final GUI. Once more, the verification of message after the interaction failed as expected in the Balsamiq prototype due to the untraceable interaction between screens. Lastly, regarding to the final test results in each artifact, we notice that only the second scenario was successfully executed in the final GUI. The other two scenarios have failed in the other artifacts or got pending (the last scenario in the final GUI).

Table 3. Results after assessing the artifacts.

| Artifact | Total (Steps Analyzed) | Results | |
|---|---|---|---|
| | | *Consistent* | *Inconsistent* |
| Task Models | 147 | 5 | 142 |
| GUI Prototypes | 36 | 21 | 15 |
| Final GUIs | 288 | 276 | 12 |

Table 3 summarizes the results obtained after multiple batteries of tests with different scenarios. For task models, the most common source of the 142 inconsistencies identified concerned the task gaps present at the beginning of the scenario. As the assessment is performed in the extracted scenarios which represent a sequential instance of the tasks in the task model, a task gap at the beginning causes a domino effect to the forthcoming tasks in the scenario. So even if the remaining tasks in the scenario are semantically equivalent to the respective steps, they will be shown as inconsistent once they will be found at wrong positions due to this gap. For GUI prototypes, from the 15 inconsistencies identified, we noticed they were mainly due to interaction elements specified with different names in the step and in the prototype. For final GUIs, the high number of consistent steps (276 out of 288) in the set of scenarios analyzed is due to the need of fixing the inconsistency found before moving forward to the next steps. This makes that the scenarios which call previous ones, in order to reuse steps and reach a given state of the system, already have these steps fully consistent during the test. Most part of the inconsistencies on final GUIs was due to interaction elements that did not carry a unique and single identifier (or carried a dynamically generated one) and, as such, could not be reached during the test.

## 5. Discussion

### *5.1. Types of Inconsistencies Identified*

Table 4. Main kinds of problems identified in each artifact after testing.

| Task Models | Balsamiq Prototypes | Final GUIs |
|---|---|---|
| • Task with different names <br> • Task not extracted to the scenario <br> • Different number of sequences of tasks in the task model <br> • Wrong position <br> • Conflict between specification and modeling <br> • Different specification strategies <br> • Unpaired behaviors <br> • Equivalent behaviors missing | • Conflict between expected and actual elements <br> • Element and label in different groups <br> • Inexistent elements <br> • Element semantically inconsistent <br> • More than one element to represent the same field <br> • Untraceable interaction between screens | • Message not identified <br> • Element or value not found <br> • Inexistent elements <br> • Values that do not fit the field <br> • Fields already filled in <br> • Element not identified |

By analyzing the variety of inconsistency problems that have been identified in this case study, we can remark that some types of inconsistencies have shown to be more critical than others. While simple inconsistencies such as differences in names of tasks and fields are easy to be solved, some other inconsistencies can reveal crucial problems of modeling or important incompatibilities between the requirements specification and its modeling in the artifacts. Conflicts between expected and actual elements in prototypes (usually due to different names), or messages and elements not found (or even inexistent) in prototypes or final GUIs are other examples of inconsistencies that are easy to solve. For simple inconsistencies such as differences in names for example, the use of a dictionary or the implementation of a feature to automatically fix the names after user confirmation could be very helpful.

Conflicts between specification and modeling along with different specification strategies for task models compose a more critical group of problems and must be prioritized. Concerning GUI prototypes, the presence of semantically inconsistent elements and the presence of more than one element to represent the same field are also critical groups of problems. On final GUIs, fields already filled-in denotes inconsistencies that exposes important design errors.

Unpaired behaviors and equivalent behaviors missing on task models are inconsistencies directly related to the extent of the referenced ontology. By evolving the ontology in the future, it can eventually support new behaviors, which would increase its capacity of recognizing other task descriptions. Tasks not extracted to the scenario, different number of task sequences in the task model, and the presence of tasks at wrong positions are problems that can have the same origin in scenarios extracted from task models, i.e. due to tasks not extracted to scenarios, the ones that have been extracted may be placed at wrong positions which will affect how many sequences of tasks the scenario lists. So, by fixing the root cause, the designer can avoid three kinds of different problems.

Element and label in different groups is an inconsistency related to the way Balsamiq implements its prototypes. It leads to a misidentification of elements but could eventually not be an issue in other prototyping tools. Likewise, elements not identified in the final GUI is a specific problem of web interfaces, where not even robust XPaths can be able to identify some elements on the interface. Values that do not fit the field is another exclusive problem of final GUIs, once it emerges from real data handling along the interaction.

Untraceable interaction between screens is a particular type of inconsistency due to the level of refinement we are considering for prototypes. Balsamiq is a prototype tool that actually supports a basic dialog description, allowing building links between prototypes and simulating a real navigation on user interfaces. However, we have chosen to not consider such a feature since it would imply a co-execution of tests on the prototypes. Such a strategy for testing has already been covered by more robust interactions in other levels of prototype refinements, such as the one that has been implemented by PANDA [49] using the same supporting ontology.

We notice finally that our approach makes task models, GUI prototypes and final GUIs intrinsically related in such a way that a modification in the BDD scenarios requires a full battery of regression tests in the other artifacts. For example, after having fixed the Balsamiq prototype to name the three-company field independently, we had to modify the step *"And I set 'Air France' in the field 'Companies'"* to reference only one of them. We modified the step to *"And I set 'AF' in the field 'Company 1'"*. When doing this, we needed to retest all our task model scenarios in order to ensure we did not introduce any inconsistency there due to such a modification. Eventually the task model has to be modified as well, in order to comply with the modification in the prototype. The modification has also the potential to identify new inconsistencies in the task model, that had not been identified before with the older scenario.

We also noticed that some identified inconsistencies were due to a wrong specification of the step in the BDD scenario, and not to a problem in the design of the artifact itself. So, to fix these inconsistencies, steps of BDD scenarios needed to be modified during the battery of tests to obtain a consistent specification of user requirements and artifacts. An immediate consequence of this fact is that scenarios used to test a given version of an artifact may be different than the ones which were used to test another artifact previously. This makes regression tests essential to ensure that a given modification in the set of BDD scenarios does not break the consistency of other artifacts and end up making some artifact (that so far was consistent with the requirements) inconsistent again, which reinforces the importance of the automated testing.

### 5.2. Limitations

**Extension of the covered artifacts**. As limitations of the approach, it is worthwhile to mention that its current version covers only the assessment of HAMSTERS task models, Balsamiq GUI prototypes and final web GUIs. Despite the approach has been built with flexibility in mind and its implementation architecture clearly delineate a core to be modified when designing tasks models and GUI prototypes using other notations and tools, we have not yet explored such possibilities so this should be covered by future works. Still regarding this matter, the need of extracting scenarios from task models to perform testing in such artifacts is another limitation. Further studies should investigate the assessment directly on task models, without necessarily extracting scenarios from them.

**Generalization of the results.** We have conducted the present study by modeling and assessing a system for booking airline tickets for business trips. Such kind of system has usually a strong search-based feature since they are centered in providing and comparing rates, times, and availability of flights given a set of provided parameters. As the ontology we use is designed to cover only domain-independent interactive behaviors, we assume that our results concerning the usefulness of our approach to identifying different inconsistencies on the considered artifacts would be reproduced and could be generalized to other search-based interactive systems domains. We acknowledge; however, the obtained results are still preliminary, produced in a specific context, in a study with some limitations. Thus, despite the results are promising so far, other studies are still necessary to allow wider conclusions.

**Manual reverse engineering**. This study performed a manual reverse engineering of the current system in production to obtain its respective models for testing. The goal of the study was to investigate the extent of inconsistencies our approach would be able to identify in the models and in the system. Therefore, as a manual process, it was expected that inconsistencies would be naturally introduced during the modeling. Indeed, these inconsistencies were identified and that allowed us to evaluate our approach. Nonetheless, if an automated approach of reverse engineering had been used instead, such inconsistencies would probably not have taken place. Future studies should confirm this hypothesis.

**Possible modeling bias**. Since both the conduction of the study and the interpretation and analysis of the results have initially been made by the authors, a possible bias should be considered. To mitigate that, the results were cross-checked by independent reviewers, experts in software engineering and modeling. They examined both the reengineered artifacts and the testing results, then they performed a qualitative analysis of the types of inconsistencies identified. The results presented in this article are thus a consolidated and revised version of the testing outcomes.

# 6. Conclusion

Artifacts provide specialized views for concepts handled by models, ensuring that aspects of the system in consideration are properly described and understood by stakeholders. Multiple cycles of design and evaluation allow to tune the design and fix problems iteratively until all requirements are met. Whereas models and iterative processes are in use, a dangling question remains: how to ensure that models and artifacts remain consistent along an iterative development process? In this context, the present work contributes by providing an approach for specifying and testing user requirements in order to keep the consistency of such requirements with core software artifacts commonly used to build interactive systems and their user interfaces.

Compared to plain-vanilla BDD, this approach contributes and benefits researchers and practitioners with (*i*) a tool-supported extension to assess user interface design artifacts prior to the development of a fully-functional software system, and (*ii*) a common vocabulary to be reused for specifying BDD scenarios without requiring developers to implement each interactive behavior for testing. Compared to other approaches for assessing requirements and artifacts, the term "test" is usually not employed under the argument that such artifacts cannot be "run", i.e. executed for testing purposes, so in practice they are just manually reviewed or inspected in a process called verification. Manual verification of the software outcomes is highly time-consuming, error-prone and even impracticable for large software systems. Fully interactive artifacts such as final GUIs can in addition be validated by users who can interact with the artifact and assess whether its behavior is aligned with their actual needs. As within our approach we succeed automatically running BDD stories on software artifacts for assessing their consistency with user requirements, we actually provide the "test" component for both verification and validation of artifacts in the software development. We consider this a big step towards the automated testing (and not only the manual verification) of software artifacts by means of a consistent approach allowing fully verification, validation, and testing (VV&T).

Future works around this theme include evaluating the impact of maintaining and successively evolving the mentioned artifacts throughout a real software development process, besides investigating the suitability of the approach for assessing a wider group of artifacts, especially those related to conceptual aspects of software modeling such as class diagrams. Concerning our toolset, the development of a plugin to suggest and autocomplete steps in the BDD scenarios based on the interactive behaviors of the ontology is also envisioned, besides providing support to the automatic classification of errors identified during the assessment.

**References**

[1] M. Winckler and P. Palanque, "Models as Representations for Supporting the Development of e-Procedures," in Usability in Government Systems, Elsevier, 2012, pp. 301–315.

[2] S. Ambler, Agile Modeling: Effective Practices for eXtreme Programming and the Unified Process, 1st ed. Wiley, 2002.

[3] D. Chelimsky, D. Astels, B. Helmkamp, D. North, Z. Dennis, and A. Hellesoy, The RSpec Book: Behaviour Driven Development with RSpec, Cucumber, and Friends. Pragmatic Bookshelf, 2010.

[4] F. Zampetti et al., "Demystifying the adoption of behavior-driven development in open source projects," Inf. Softw. Technol., vol. 123, no. July, 2020, doi: 10.1016/j.infsof.2020.106311.

[5] L. Pereira, H. Sharp, C. De Souza, G. Oliveira, S. Marczak, and R. Bastos, "Behavior-driven development benefits and challenges: Reports from an industrial study," in XP '18: Proceedings of the 19th International Conference on Agile Software Development: Companion, 2018, vol. Article No, pp. 1–4, doi: 10.1145/3234152.3234167.

[6] M. Cohn, User Stories Applied for Agile Software Development. Addison-Wesley, 2004.

[7] T. R. Silva, M. Winckler, and H. Trætteberg, "Extending Behavior-Driven Development for Assessing User Interface Design Artifacts," in Proceedings of the International Conference on Software Engineering and Knowledge Engineering, SEKE, 2019, pp. 485–488, doi: 10.18293/SEKE2019-054.

[8] J. F. Smart, BDD in Action: Behavior-driven development for the whole software lifecycle, 1 edition. Manning Publications, 2014.

[9] D. North, "What's in a Story?," 2019. [Online]. Available: https://dannorth.net/whats-in-a-story/. [Accessed: 01-Jan-2019].

[10] J. H. Lopes, "Evaluation of Behavior-Driven Development," Delft University of Technology, 2012.

[11] C. Solís and X. Wang, "A Study of the Characteristics of Behaviour Driven Development," in Proceedings - 37th EUROMICRO Conference on Software Engineering and Advanced Applications, SEAA 2011, 2011, pp. 383–387, doi: 10.1109/SEAA.2011.76.

[12] A. Egbreghts, "A Literature Review of Behavior Driven Development using Grounded Theory," in 27th Twente Student Conference on IT, 2017.

[13] G. I. Melnik, "Empirical Analyses of Executable Acceptance Test Driven Development," University of Calgary, 2007.

[14] P. Valente, T. Silva, M. Winckler, and N. Nunes, "The Goals Approach: Agile Enterprise Driven Software Development," in Complexity in Information Systems Development, vol. 22, 2017, pp. 201–219.

[15] P. Valente, T. Silva, M. Winckler, and N. Nunes, "Bridging Enterprise and Software Engineering Through an User-Centered Design Perspective," in Web Information Systems Engineering – WISE 2016, 2016, vol. 10042 LNCS, pp. 349–357, doi: 10.1007/978-3-319-48743-4.

[16] P. Lombriser, F. Dalpiaz, G. Lucassen, and S. Brinkkemper, "Gamified Requirements Engineering: Model and Experimentation," in Proceedings of the 22nd International Working Conference on Requirements Engineering: Foundation for Software Quality (REFSQ 2016), 2016, pp. 171–187.

[17] M. Rahman and J. Gao, "A Reusable Automated Acceptance Testing Architecture for Microservices in Behavior-Driven Development," in Proceedings - 9th IEEE International Symposium on Service-Oriented System Engineering, IEEE SOSE 2015, 2015, vol. 30, pp. 321–325, doi: 10.1109/SOSE.2015.55.

[18] A. C. Oran, E. Nascimento, G. Santos, and T. Conte, "Analysing Requirements Communication Using Use Case Specification and User Stories," in Proceedings of the 31st Brazilian Symposium on

Software Engineering (SBES 2017), 2017, pp. 214–223, doi: https://doi.org/10.1145/3131151.3131166.

[19] Y. Wang and S. Wagner, "Combining STPA and BDD for Safety Analysis and Verification in Agile Development: A Controlled Experiment," in International Conference on Agile Software Development (XP 2018), 2018, pp. 37–53, doi: https://doi.org/10.1007/978-3-319-91602-6_3.

[20] S.-T. Lai, F.-Y. Leu, and W. C.-C. Chu, "Combining IID with BDD to Enhance the Critical Quality of Security Functional Requirements," in 2014 Ninth International Conference on Broadband and Wireless Computing, Communication and Applications (BWCCA), 2014.

[21] G. Lucassen, F. Dalpiaz, J. M. E. M. Van Der Werf, S. Brinkkemper, and Di. Zowghi, "Behavior-Driven Requirements Traceability via Automated Acceptance Tests," in Proceedings - 2017 IEEE 25th International Requirements Engineering Conference Workshops, REW 2017, 2017, pp. 431–434, doi: 10.1109/REW.2017.84.

[22] A. Z. H. Yang, D. Alencar Da Costa, and Y. Zou, "Predicting co-changes between functionality specifications and source code in behavior driven development," in IEEE/ACM 16th International Conference on Mining Software Repositories (MSR), 2019, vol. 2019-May, pp. 534–544, doi: 10.1109/MSR.2019.00080.

[23] R. A. de Carvalho, F. L. de Carvalho e Silva, and R. S. Manhaes, "Mapping Business Process Modeling constructs to Behavior Driven Development Ubiquitous Language," 2010.

[24] R. A. de Carvalho, R. S. Manhães, and F. L. de Carvalho e Silva, "Filling the Gap between Business Process Modeling and Behavior Driven Development," 2010.

[25] D. Lübke and T. Van Lessen, "Modeling Test Cases in BPMN for Behavior- Driven Development," IEEE Softw., no. October, pp. 15–21, 2016, doi: 10.1109/MS.2016.117.

[26] M. Soeken, R. Wille, and R. Drechsler, "Assisted Behavior Driven Development Using Natural Language Processing," in TOOLS Europe 2012, 2012, vol. 7304 LNCS, pp. 269–287.

[27] F. Paternò, C. Santoro, L. D. Spano, and D. Raggett, "W3C, MBUI - Task Models," 2017. [Online]. Available: http://www.w3.org/TR/task-models/.

[28] C. Martinie, P. Palanque, and M. A. Winckler, "Structuring and Composition Mechanisms to Address Scalability Issues in Task Models," in Proc. of the IFIP TC.13 International Conference on Human-Computer Interaction (INTERACT), 2011, vol. 6948 LNCS, no. Part 3, pp. 589–609, doi: 10.1007/978-3-642-23765-2_40.

[29] M. Beaudouin-Lafon and W. E. Mackay, "Prototyping Tools and Techniques," in Prototype Development and Tools, 2000, pp. 1–41.

[30] O. Laitenberger and J.-M. DeBaud, "An encompassing life cycle centric survey of software inspection," J. Syst. Softw., vol. 50, no. 1, pp. 5–31, 2000, doi: 10.1016/S0164-1212(99)00073-4.

[31] D. L. Parnas and M. Lawford, "The Role of Inspection in Software Quality Assurance," IEEE Trans. Softw. Eng., vol. 29, no. 8, pp. 674–676, 2003, doi: 10.1109/TSE.2003.1223642.

[32] C. Ebert, Global Software and IT: A Guide to Distributed Development, Projects, and Outsourcing. John Wiley & Sons, 2011.

[33] E. R. Luna, I. Garrigós, J. Grigera, and M. Winckler, "Capture and Evolution of Web Requirements Using WebSpec," in Proc. of the Int. Conference on Web Engineering, 2010, vol. 6189 LNCS, pp. 173–188, doi: 10.1007/978-3-642-13911-6_12.

[34] R. A. Buchmann and D. Karagiannis, "Modelling mobile app requirements for semantic traceability," Requir. Eng., vol. 22, no. 1, pp. 41–75, 2017, doi: 10.1007/s00766-015-0235-1.

[35] M. Elkoutbi, I. Khriss, and R. K. Keller, "Generating user interface prototypes from scenarios," in Proceedings IEEE International Symposium on Requirements Engineering (Cat. No.PR00188), 1999, pp. 150–158, doi: 10.1109/ISRE.1999.777995.

[36] I. Khaddam, N. Mezhoudi, and J. Vanderdonckt, "Towards Task-Based Linguistic Modeling for Designing GUIs," in 27th Conference on l'Interaction Homme-Machine, 2015.

[37] C. Martinie, P. Palanque, and M. Winckler, "Designing and Assessing Interactive Systems Using Task Models," in IFIP Conference on Human-Computer Interaction, 2015.

[38] J. C. Campos, C. Fayollas, C. Martinie, D. Navarre, P. Palanque, and M. Pinto, "Systematic Automation of Scenario-Based Testing of User Interfaces," in Proceedings of the 8th ACM SIGCHI Symposium on Engineering Interactive Computing Systems - EICS '16, 2016, pp. 138–148, doi: 10.1145/2933242.2948735.

[39] T. R. Silva, J.-L. Hak, and M. Winckler, "A Behavior-Based Ontology for Supporting Automated Assessment of Interactive Systems," in Proceedings - IEEE 11th International Conference on Semantic Computing, ICSC 2017, 2017, pp. 250–257, doi: 10.1109/ICSC.2017.73.

[40] T. R. Silva, J.-L. Hak, and M. Winckler, "A Formal Ontology for Describing Interactive Behaviors and Supporting Automated Testing on User Interfaces," Int. J. Semant. Comput., vol. 11, no. 04, pp. 513–539, 2017, doi: 10.1142/S1793351X17400219.

[41] G. Calvary et al., "The CAMELEON Reference Framework," 2002.

[42] Q. Limbourg, J. Vanderdonckt, B. Michotte, L. Bouillon, and V. López-Jaquero, "USIXML: A Language Supporting Multi-path Development of User Interfaces," in EHCI/DS-VIS, 2004, no. LNCS 3425, pp. 200–220, doi: 10.1007/11431879_12.

[43] J. Pullmann, "MBUI - Glossary - W3C," 2017. [Online]. Available: https://www.w3.org/TR/mbui-glossary/. [Accessed: 01-Dec-2017].

[44] M. Winckler and P. Palanque, "StateWebCharts: A Formal Description Technique Dedicated to Navigation Modelling of Web Applications," in Interactive Systems. Design, Specification, and Verification, 2003, pp. 61–76, doi: 10.1007/978-3-540-39929-2_5.

[45] F. Paternò, "ConcurTaskTrees: An Engineered Notation for Task Model," in The Handbook of Task Analysis for Human-Computer Interaction, Lawrence Erlbaum Associates, 2003, pp. 483–503.

[46] T. R. Silva, J.-L. Hak, and M. A. Winckler, "A Review of Milestones in the History of GUI Prototyping Tools," in IFIP TC.13 International Conference on Human-Computer Interaction – INTERACT 2015 Adjunct Proceedings, 2015, pp. 1–13.

[47] T. R. Silva, J.-L. Hak, M. Winckler, and O. Nicolas, "A Comparative Study of Milestones for Featuring GUI Prototyping Tools," J. Softw. Eng. Appl., vol. 10, no. 06, pp. 564–589, 2017, doi: 10.4236/jsea.2017.106031.

[48] T. Rocha Silva, M. Winckler, C. Bach, T. R. Silva, M. Winckler, and C. Bach, "Evaluating the usage of predefined interactive behaviors for writing user stories: an empirical study with potential product owners," Cogn. Technol. Work, May 2019, doi: 10.1007/s10111-019-00566-3.

[49] J. Hak, M. Winckler, and D. Navarre, "PANDA : Prototyping using ANnotation and Decision Analysis," in Proceedings of the 8th ACM SIGCHI Symposium on Engineering Interactive Computing Systems, 2016, pp. 171–176.

# Appendix A: Concept Mapping Table

| Checkbox and Radio Button Behaviors | | | | | |
|---|---|---|---|---|---|
| | | | **Interaction Elements Affected** | | |
| **Interactive Behavior** | **Task** | **Step of Scenario** | **Ontology** | **Balsamiq Prototype** (com.balsamiq.mockups::) | **Final GUI** |
| *theFieldIsUnchecked* | *Verify the field <fieldname> is unchecked* | **Given/When/Then** *the field "<fieldname>" is unchecked* | Checkbox | CheckBox | CheckBox |
| | | | Radio Button | RadioButton | Radio |
| *theFieldIsChecked* | *Verify the field <fieldname> is checked* | **Given/When/Then** *the field "<fieldname>" is checked* | Checkbox | CheckBox | CheckBox |
| | | | Radio Button | RadioButton | Radio |
| *assureTheFieldIsUnchecked* | *Assure the field <fieldname> is unchecked* | **When** *I assure the field "<fieldname>" is unchecked* | Checkbox | CheckBox | CheckBox |
| *assureTheFieldIsChecked* | *Assure the field <fieldname> is checked* | **When** *I assure the field "<fieldname>" is checked* | Checkbox | CheckBox | CheckBox |
| **Common Behaviors** | | | | | |
| | | | **Interaction Elements Affected** | | |
| **Interactive Behavior** | **Task** | **Step of Scenario** | **Ontology** | **Balsamiq Prototype** (com.balsamiq.mockups::) | **Final GUI** |
| *choose* | *Choose <option>* | **Given/When/Then** *I choose "<option>"* | Calendar | Calendar or DateChooser | Calendar |
| | | | Checkbox | CheckBox | CheckBox |
| | | | Radio Button | RadioButton | Radio |
| | | | Link | Link | Link |
| *select* | *Select <option>* | **Given/When/Then** *I select "<option>"* | Calendar | Calendar or DateChooser | Calendar |
| | | | Checkbox | CheckBox | CheckBox |
| | | | Radio Button | RadioButton | Radio |
| | | | Link | Link | Link |
| *chooseByIndexInTheField* | *Choose in the field <fieldname>* | **When/Then** *I choose "<index>" by index in the field "<fieldname>"* | Dropdown List | ComboBox | Select |
| *chooseReferringTo* | *Choose <fieldname> referring to <option>* | **When/Then** *I choose "<fieldname>" referring to "<option>"* | Calendar | Calendar or DateChooser | Calendar |
| | | | Checkbox | CheckBox | CheckBox |
| | *Choose <fieldname>* | | Radio Button | RadioButton | Radio |
| | | | Link | Link | Link |

| | | | Interaction Elements Affected | | |
|---|---|---|---|---|---|
| **Interactive Behavior** | **Task** | **Step of Scenario** | **Ontology** | **Balsamiq Prototype** (com.balsamiq.mockups::) | **Final GUI** |
| *chooseTheOptionOfValueInThe Field* | *Choose in the field <fieldname>* | **When/Then** *I choose the option of value "<value>" in the field "<fieldname>"* | Dropdown List | ComboBox | Select |
| *clickOn* | *Click on <fieldname>* | **When/Then** *I click on "<fieldname>"* | Menu | MenuBar | Menu |
| | | | Menu Item | Accordion | MenuItem |
| | | | Button | Button | Button |
| | | | Link | Link | Link |
| *clickOnReferringTo* | *Click on <fieldname> referring to <option>* | **When/Then** *I click on "<fieldname>" referring to "<option>"* | Menu | MenuBar | Menu |
| | | | Menu Item | Accordion | MenuItem |
| | | | Button | Button | Button |
| | | | Link | Link | Link |
| | *Click on <fieldname>* | | Grid | DataGrid | Grid |
| *doNotTypeAnyValueToTheField* | *Do not type any value to the field <fieldname>* | **When** *I do not type any value to the field "<fieldname>"* | Text Field | TextInput | TextField |
| *resetTheValueOfTheField* | *Reset the value of the field <fieldname>* | **When** *I reset the value of the field "<fieldname>"* | Text Field | TextInput | TextField |
| *goTo* | *Go to <address>* | **Given/When/Then** *I go to "<address>"* | Browser Window | BrowserWindow | Screen |
| *goToWithTheParameters* | *Go to <address> with the parameters <parameters>* | **Given/When/Then** *I go to "<address>" with the parameters "<parameters>"* | Browser Window | BrowserWindow | Screen |
| *isDisplayed* | *Display <page>* | **Given/When/Then** *"<page>" is displayed* | Browser Window | BrowserWindow | Screen |
| *setInTheField* | *Set <fieldname>* | **When/Then** *I set "<value>" in the field "<fieldname>"* | Dropdown List | ComboBox | Select |
| | | | Text Field | TextInput | TextField |
| | | | Autocomplete | SearchBox | AutoComplete |
| | | | Calendar | Calendar or DateChooser | Calendar |
| *tryToSetInTheField* | *Try to set <fieldname>* | **When/Then** *I try to set in the field "<fieldname>"* | Dropdown List | ComboBox | Select |
| | | | Text Field | TextInput | TextField |
| | | | Autocomplete | SearchBox | AutoComplete |
| | | | Calendar | Calendar or DateChooser | Calendar |
| *setInTheFieldReferringTo* | *Set <fieldname>* | | Dropdown List | ComboBox | Select |

Checkbox and Radio Button Behaviors

| | | | Interaction Elements Affected | | |
|---|---|---|---|---|---|
| **Interactive Behavior** | **Task** | **Step of Scenario** | **Ontology** | **Balsamiq Prototype** (com.balsamiq.mockups::) | **Final GUI** |
| | | **When/Then** I set "<value>" in the field referring to "fieldname>" | Text Field | TextInput | TextField |
| *typeAndChooseInTheField* | *Inform <value 1>* *Choose <value 2>* | **When/Then** I type "<value 1>" and choose "<value 2>" in the field "<fieldname>" | Autocomplete | SearchBox | AutoComplete |
| *informAndChooseInTheField* | *Inform <value 1>* *Choose <value 2>* | **When/Then** I inform "<value 1>" and choose "<value 2>" in the field "<fieldname>" | Autocomplete | SearchBox | AutoComplete |
| *willBeDisplayed* | *Display <content>* | **Then** "<content>" will be displayed | Text | Paragraph | Text |
| *willNotBeDisplayed* | *Not display <content>* | **Then** "<content>" will not be displayed | Text | Paragraph | Text |
| *willBeDisplayedInTheFieldTheValue* | *Display <value>* | **Then** will be displayed in the field "<fieldname>" the value "<value>" | Element | UI Element | Element |
| *willNotBeDisplayedInTheFieldTheValue* | *Not display <value>* | **Then** will not be displayed in the field "<fieldname>" the value "<value>" | Element | UI Element | Element |
| *willBeDisplayedTheValueInTheFieldReferringTo* | *Display <value>* | **Then** will be displayed the value "<value>" in the field "<fieldname>" referring to "<element>" | Element | UI Element | Element |
| *willNotBeDisplayedTheValueInTheFieldReferringTo* | *Not display <value>* | **Then** will not be displayed the value "<value>" in the field "<fieldname>" referring to "<element>" | Element | UI Element | Element |
| *isNotVisible* | *Hidden <fieldname>* | **Given/When/Then** "<fieldname>" is not visible | Element | UI Element | Element |
| *valueReferringToIsNotVisible* | *Hidden <value>* | **Given/When/Then** "<value>" referring to "<element>" is not visible | Element | UI Element | Element |
| *waitTheFieldBeVisibleClickableAndEnable* | *Wait the field <fieldname> be visible, clickable and enable* | **Given/When/Then** I wait the field "<fieldname>" be visible, clickable and enable | Element | UI Element | Element |
| *waitTheFieldReferringToBeVisibleClickableAndEnable* | *Wait the field <fieldname> be visible, clickable and enable* | **Given/When/Then** I wait the field "<fieldname>" referring to "<element>" be visible, clickable and enable | Element | UI Element | Element |
| *theElementIsVisibleAndDisable* | *Check the element <element> is visible and disable* | **Given/When/Then** the element "<element>" is visible and disable | Element | UI Element | Element |
| *theElementReferringToIsVisibleAndDisable* | *Check the element <element> is visible and disable* | **Given/When/Then** the element "<fieldname>" referring to "<element>" is visible and disable | Element | UI Element | Element |
| *setInTheFieldAndTriggerTheEvent* | *Set <fieldname>* *Trigger <event>* | **When/Then** I set in the field "<fieldname>" and trigger the event "<event>" | Text Field | TextInput | TextField |

Checkbox and Radio Button Behaviors

## Checkbox and Radio Button Behaviors

| Interactive Behavior | Task | Step of Scenario | Interaction Elements Affected | | |
|---|---|---|---|---|---|
| | | | **Ontology** | **Balsamiq Prototype** (com.balsamiq.mockups::) | **Final GUI** |
| *clickOnTheRowOfTheTree* | Select value for <tree> | **Given/When/Then** I click on the row "<row>" of the tree "<tree>" | Tree | - | Tree |

## Data Generation Behaviors

| Interactive Behavior | Task | Step of Scenario | Interaction Elements Affected | | |
|---|---|---|---|---|---|
| | | | **Ontology** | **Balsamiq Prototype** (com.balsamiq.mockups::) | **Final GUI** |
| *informARandomNumberWithPrefixInTheField* | Inform a random number with prefix in the field <fieldname> | **Given/When/Then** I inform a random number with prefix "<prefix>" in the field "<fieldname>" | Text Field | TextInput | TextField |
| *informARandomNumberInTheField* | Inform a random number in the field <fieldname> | **When** I inform a random number in the field "<fieldname>" | Text Field | TextInput | TextField |

## Data Provider Behaviors

| Interactive Behavior | Task | Step of Scenario | Interaction Elements Affected | | |
|---|---|---|---|---|---|
| | | | **Ontology** | **Balsamiq Prototype** (com.balsamiq.mockups::) | **Final GUI** |
| *inform* | Inform <value> | **Given/When** I inform "<value>" | Grid | DataGrid | Grid |
| *informTheField* | Inform the field <fieldname> | **When** I inform the field "<fieldname>" | Grid | DataGrid | Grid |
| *informTheFields* | Inform the fields <fieldnames> | **When** I inform the fields "<fieldnames>" | Grid | DataGrid | Grid |
| *selectFromDataSet* | Select from dataset <dataset> | **Given/When** I select from dataset "<dataset>" | - | - | - |
| *informTheValueOfTheField* | Inform the value of the field <fieldname> | **When/Then** I inform the value of the field "<fieldname>" | Element | UI Element | Element |
| *informKeyWithTheValue* | Inform key <key> | **Given/When/Then** I inform key "<key>" with the value "<value>" | - | - | - |
| *defineTheVariableWithTheValue* | Define the variable <variable> | **Given/When/Then** I define the variable "<variable>" with the value "<value>" | - | - | - |
| *obtainTheValueFromTheField* | Obtain the value from the field <fieldname> | **Given/When/Then** I obtain the value from the field "<fieldname>" | Element | UI Element | Element |

## Debug Behaviors

| Interactive Behavior | Task | Step of Scenario | Interaction Elements Affected | | |
|---|---|---|---|---|---|
| | | | **Ontology** | **Balsamiq Prototype** | **Final GUI** |

| Checkbox and Radio Button Behaviors | | | | | |
|---|---|---|---|---|---|
| **Interactive Behavior** | **Task** | **Step of Scenario** | **Interaction Elements Affected** | | |
| | | | **Ontology** | **Balsamiq Prototype** (com.balsamiq.mockups::) | **Final GUI** |
| *printOnTheConsoleTheValueOfTheVariable* | *Print on the console the value of the variable <variable>* | **When/Then** *I print on the console the value of the variable <variable>* | - | - | - |
| Dialog Behaviors | | | | | |
| **Interactive Behavior** | **Task** | **Step of Scenario** | **Interaction Elements Affected** | | |
| | | | **Ontology** | **Balsamiq Prototype** (com.balsamiq.mockups::) | **Final GUI** |
| *confirmTheDialogBox* | *Confirm the dialog box* | **Given/When/Then** *I confirm the dialog box* | Window Dialog | Alert | Dialog |
| *cancelTheDialogBox* | *Cancel the dialog box* | **Given/When/Then** *I cancel the dialog box* | Window Dialog | Alert | Dialog |
| *informTheValueInTheDialogBox* | *Inform the value in the dialog box* | **Given/When/Then** *I inform the value "<value>" in the dialog box* | Window Dialog | Alert | Dialog |
| *willBeDisplayedInTheDialogBox* | *Display <message> in the dialog box* | **Then** *will be displayed "<message>" in the dialog box* | Window Dialog | Alert | Dialog |
| Mouse Control Behaviors | | | | | |
| **Interactive Behavior** | **Task** | **Step of Scenario** | **Interaction Elements Affected** | | |
| | | | **Ontology** | **Balsamiq Prototype** (com.balsamiq.mockups::) | **Final GUI** |
| *moveTheMouseOver* | *Move the mouse over <element>* | **When** *I move the mouse over "<element>"* | Menu | MenuBar | Menu |
| | | | Menu Item | Accordion | MenuItem |
| | | | Button | Button | Button |
| | | | Link | Link | Link |
| Table Behaviors | | | | | |
| **Interactive Behavior** | **Task** | **Step of Scenario** | **Interaction Elements Affected** | | |
| | | | **Ontology** | **Balsamiq Prototype** (com.balsamiq.mockups::) | **Final GUI** |
| *clickOnTheRowOfTheTableReferringTo* | *Click on the row of the table <table>* | **When/Then** *I click on the row "<row>" of the table "<table>" referring to "<element>"* | Grid | DataGrid | Grid |
| *storeTheCellOfTheTableIn* | *Store the cell of the table <table> in <place>* | **When/Then** *I store the cell "<cell>" of the table "<table>" in "<place>"* | Grid | DataGrid | Grid |

| | | | Interaction Elements Affected | | |
|---|---|---|---|---|---|
| **Checkbox and Radio Button Behaviors** | | | | | |
| **Interactive Behavior** | **Task** | **Step of Scenario** | **Ontology** | **Balsamiq Prototype** (com.balsamiq.mockups::) | **Final GUI** |
| *storeTheColumnOfTheTableIn* | *Store the column of the table <table> in <place>* | **When/Then** *I store the column "<column>" of the table "<table>" in "<place>"* | Grid | DataGrid | Grid |
| *compareTheTextOfTheTableCellWith* | *Compare the text of the table cell with <text>* | **When/Then** *I compare the text of the table cell "<table text>" with "<text>"* | Grid | DataGrid | Grid |
| *compareTheTextOfTheTableColumnWith* | *Compare the text of the table column with <text>* | **When/Then** *I compare the text of the table column "<table text>" with "<text>"* | Grid | DataGrid | Grid |
| *clickOnTheCellOfTheTable* | *Click on the cell of the table <table>* | **When/Then** *I click on the cell "<cell>" of the table "<table>"* | Grid | DataGrid | Grid |
| *clickOnTheColumnOfTheTable* | *Click on the column of the table <table>* | **When/Then** *I click on the column "<column>" of the table "<table>"* | Grid | DataGrid | Grid |
| *chooseTheOptionInTheCellOfTheTable* | *Choose the option in the cell of the table <table>* | **When/Then** *I choose the option "<option>" in the cell of the table "<table>"* | Grid | DataGrid | Grid |
| *chooseTheOptionInTheColumnOfTheTable* | *Choose the option in the column of the table <table>* | **When/Then** *I choose the option "<option>" in the column of the table "<table>"* | Grid | DataGrid | Grid |
| *typeTheTextInTheCellOfTheTable* | *Type the text in the cell of the table <table>* | **When/Then** *I type the text "<text>" in the cell of the table "<table>"* | Grid | DataGrid | Grid |
| *typeTheTextInTheColumnOfTheTable* | *Type the text in the column of the table <table>* | **When/Then** *I type the text "<text>" in the column of the table "<table>"* | Grid | DataGrid | Grid |